%% file: irbol.astroph.tex
\documentclass[useAMS,usenatbib]{mn2e}

\input{epsf}

\usepackage{graphicx}	
\usepackage{deluxetable}
\usepackage{amsmath}

\input{commands.tex}

\title[Infrared Bolometric Corrections]{Updating quasar bolometric luminosity corrections. II. Infrared bolometric corrections}
\author[J. C. Runnoe et al.]{Jessie C. Runnoe$^{1}$\thanks{E-mail:
jrunnoe@uwyo.edu} , Michael S. Brotherton$^{1}$, and Zhaohui Shang$^{2}$ \\
$^{1}$Department of Physics and Astronomy, University of Wyoming, Laramie, WY 82071, USA\\
$^{2}$Department of Physics, Tianjin Normal University, Tianjin 300074, China}
\begin{document}		

\date{Preprint 2012 April 28}

\pagerange{\pageref{firstpage}--\pageref{lastpage}} \pubyear{2012}

\maketitle

\label{firstpage}

\begin{abstract}
We present infrared bolometric luminosity corrections derived from the detailed spectral energy distributions of 62 bright quasars of low- to moderate-redshift ($z=0.03-1.4$).  At 1.5, 2, 3, 7, 12, 15, and 24 $\mu$m we provide bolometric corrections of the mathematical forms $L_{iso}=\zeta\, \lambda L_{\lambda}$ and log$(L_{iso})=A+B$ log$(\lambda L_{\lambda})$.  Bolometric corrections for radio-loud and radio-quiet objects are consistent within 95\% confidence intervals, so we do not separate them.  Bolometric luminosities estimated using these corrections are typically smaller than those derived from some commonly used in the literature.  We investigate the possibility of a luminosity dependent bolometric correction and find that, while the data are consistent with such a correction, the dispersion is too large and the luminosity range too small to warrant such a detailed interpretation.  Bolometric corrections at 1.5 $\mu$m are appropriate for objects with properties that fall in the range log(L$_{bol})=45.4-47.3$ and bolometric corrections at all other wavelengths are appropriate for objects with properties that fall in the range log(L$_{bol})=45.1-47.0$.
\end{abstract}


\section{introduction}
Recently in \citet{runnoe12a}, we derived optical, ultra-violet (UV), and X-ray bolometric luminosity corrections from the quasar spectral energy distributions (SEDs) of the \citet{shang11} atlas.  We addressed the practical difficulties of determining bolometric corrections, including source variability, contamination by emission from the host galaxy, gaps in data coverage, and the subtleties of assuming isotropy when integrating the SED, and derived bolometric corrections at 1450, 3000, 5100 \AA, and $2-10$ keV.  The bolometric corrections we provided are available with zero and nonzero intercepts, are appropriate for radio-loud (RL) and radio-quiet (RQ) quasars, and come with an additional correction for viewing angle that is not built in.  Accretion discs do not emit isotropically; we differentiated between the bolometric luminosity calculated under the assumption of isotropy and the true bolometric luminosity which motivated the additional correction to be applied after their optical/UV or X-ray bolometric corrections.

In April of 2011, the first public data release from the {\it Wide-field Infrared Survey Explorer} \citep[WISE;][]{wright10} opened wide a window into the infrared (IR) of active galactic nuclei (AGN) that was most recently explored by the {\it Spitzer Space Telescope} \citep{werner04}.  The population of IR-selected AGN that is revealed by these missions, often reddened quasars, may be unavailable at optical wavelengths due to obscuration and therefore cannot use the optical/UV bolometric corrections of \citet{runnoe12a} as a viable method for determining bolometric luminosity.

There are infrared bolometric corrections available in the literature.  \citet{elvis94}, the  original SED atlas, provides a bolometric correction at 1.5 $\mu$m, but the SEDs are not as detailed and up-to-date as those of \citet{shang11}, especially in the infrared.  \citet{richards06} employs a much larger sample size in exchange for less complete data coverage and provides a bolometric correction at 3 $\mu$m.  Both of these corrections are subject to the effects of double counting in the infrared.  \citet{hopkins07} derives a double power law correction at 15 $\mu$m that is broadly consistent with \citet{richards06} from an observation-based model SED.  
	
With the release of WISE and the ensuing level of interest in the infrared emission of quasars, there is a need for new infrared bolometric corrections.  We determine infrared bolometric corrections from the SEDs of \citet{shang11} using the same methodology as in \citet{runnoe12a}.

This paper is organized as follows.  Section 2 describes the data and the \citet{shang11} SEDs.  In Section 3 we derive bolometric corrections and in Section 4 we discuss them in the context of previous work and recommend the best bolometric corrections for various situations.  Section 5 summarizes this investigation.

We adopt a cosmology with $H_0 = 70$ km s$^{-1}$ Mpc$^{-1}$, $\Omega_{\Lambda} = 0.7$, and $\Omega_{m} = 0.3$.  Note that we also differentiate between the bolometric luminosity calculated under the assumption of isotropy $(L_{iso})$ and true bolometric luminosity ($L_{bol}$), which likely differ.

\section{Sample, data, and measurements}
\label{sec:data}

Our empirical bolometric corrections are derived from the infrared-to-X-ray continua of a subset of the \citet{shang11} SEDs.  We give a brief summary of the SED sample, data, and bolometric luminosities, which are discussed by \citet{shang11} and \citet{runnoe12a} in detail.  We then present measurements of monochromatic infrared luminosities from the SEDs.

\subsection{The SED sample}
The atlas has a total of 85 objects from three different subsamples.  The `PGX' subsample consists of 22 of 23 Palomar-Green (PG) quasars in the complete sample selected by \citet{laor94,laor97} to study the soft-X-ray regime.  This subsample is UV bright and has  $z \le 0.4$.  The `FUSE-HST' subsample has 24 objects, 17 of which come from the Far Ultraviolet Spectroscopic Explorer (\emph{FUSE}) AGN program \citep{kriss01}.  This is a heterogeneous, UV-bright sample with $z<0.5$.  The `RLQ' subsample includes nearly 50 quasars originally assembled to study orientation; all members of the sample have similar extended radio luminosity which is thought to be isotropic.  The blazars originally included in this sample are excluded in the SED atlas because of their variability due to synchrotron emission from a beamed jet.  Using the method of \citet{massaro11} and data available from WISE, we verify that none of the remaining objects have a large non-thermal contribution to emission in the infrared.

Infrared coverage comes from {\it Spitzer} and near-infrared coverage is from 2 Micron All Sky Survey \citep[2MASS;][]{skrutskie06} photometry. Mid-infrared spectroscopy was obtained for 46 objects from the Spitzer Infrared Spectrograph \citep[IRS][]{houck04}.  The infrared spectra cover about $3-35$ $\mu$m rest frame for the redshifts of the sample.  Far-infrared photometry at 24, 70, and 160 $\mu$m from Multiband Imaging Photometer on {\it Spitzer} \citep[MIPS;][]{rieke04} is available for 50 objects, all using photometry mode.  

AGN variability, which can be a concern, does not appear to be a significant issue in these SEDs.  The data in the infrared through X-ray region of the SED were collected between 1991 and 2007.  Optical-FUV data were taken quasi-simultaneously (within weeks) and infrared emission, which arises from a size scale on the order of parsecs \citep[e.g.,][]{raban09} or less for the hottest dust, will likely vary on a longer timescale of months to years.  There are no strong discontinuities in the infrared emission that might indicate variability, but we investigate it in detail.

WISE data are available for 34 objects in our sample that have IRS spectra from {\it Spitzer}, allowing us to determine infrared variability on the timescale between the {\it Spitzer} and WISE missions.  IRS coverage is such that we are only able to make this comparison in the W3 and W4 bands, at 12 and 22~$\mu$m respectively.  We use the response curves and magnitude zero points for WISE from \citet{jarrett11} to calculate WISE magnitudes from our {\it Spitzer} spectra and compare these to the observed WISE magnitudes downloaded from the WISE All-Sky Data Release.  We apply a flux correction to fix a known discrepancy that arises between red and blue calibrator stars, described in detail in \citet{wright10}, to the WISE All-Sky Data Release magnitudes of $+17\%$ and $-9\%$ in flux to the W3 and W4 bands respectively.  The resulting WISE$_{{\it Spitzer}}$ and WISE$_{\textrm{All-Sky}}$ magnitudes are given in Table~\ref{tab:spitzerwise}.  

\input{irvar.tex}

The narrow distribution of differences between WISE and {\it Spitzer} magnitudes indicates that these objects have varied by less than $5-10\%$ during the $4-5$ years between the two missions (these {\it Spitzer} data were taken in Cycle 2, which began in June 2005 and lasted one year, and WISE data were taken during the first half of 2010).  The obvious exception is 4C 11.69, which is highlighted in Table~\ref{tab:spitzerwise}.  This is a core-dominated object with log $R=1.47$ that likely has a significant synchrotron contribution to emission in the infrared that was not classified as a blazar based on optical variability by \citet{shang11} or infrared colors using the method of \citet{massaro11}.  The strong infrared variability is clear evidence that this object is a blazar that was missed by other identification methods so we exclude this object from the sample for the rest of our analysis.

\citet{shang11} applied two important corrections to the data while assembling the SEDs.  First, at optical-to-far-ultraviolet wavelengths the SEDs suffer from Galactic extinction from dust.  This is removed with the empirical mean extinction law of \citet{ccm89} using the dust maps of \citet{schlegel98}.  Second, \citet{shang11} corrects for host galaxy contamination at near infrared and optical wavelengths, although host galaxy contamination to the AGN light is not usually large for their UV/optically bright quasars it can be significant even in higher luminosity quasars \citep{lusso10}.  

Note that as a result of correcting the 2MASS points for host contamination, the shape of our SEDs will be different than an uncorrected SED in that region.  Host fractions measured by \citet{shang11} in the J-band had the range $0.08-0.94$, with an average of $0.40$.  In some objects this can be a significant effect, more so at low luminosities.

\begin{figure*}
\begin{center}
\includegraphics[width=13.0 truecm, angle=-90]{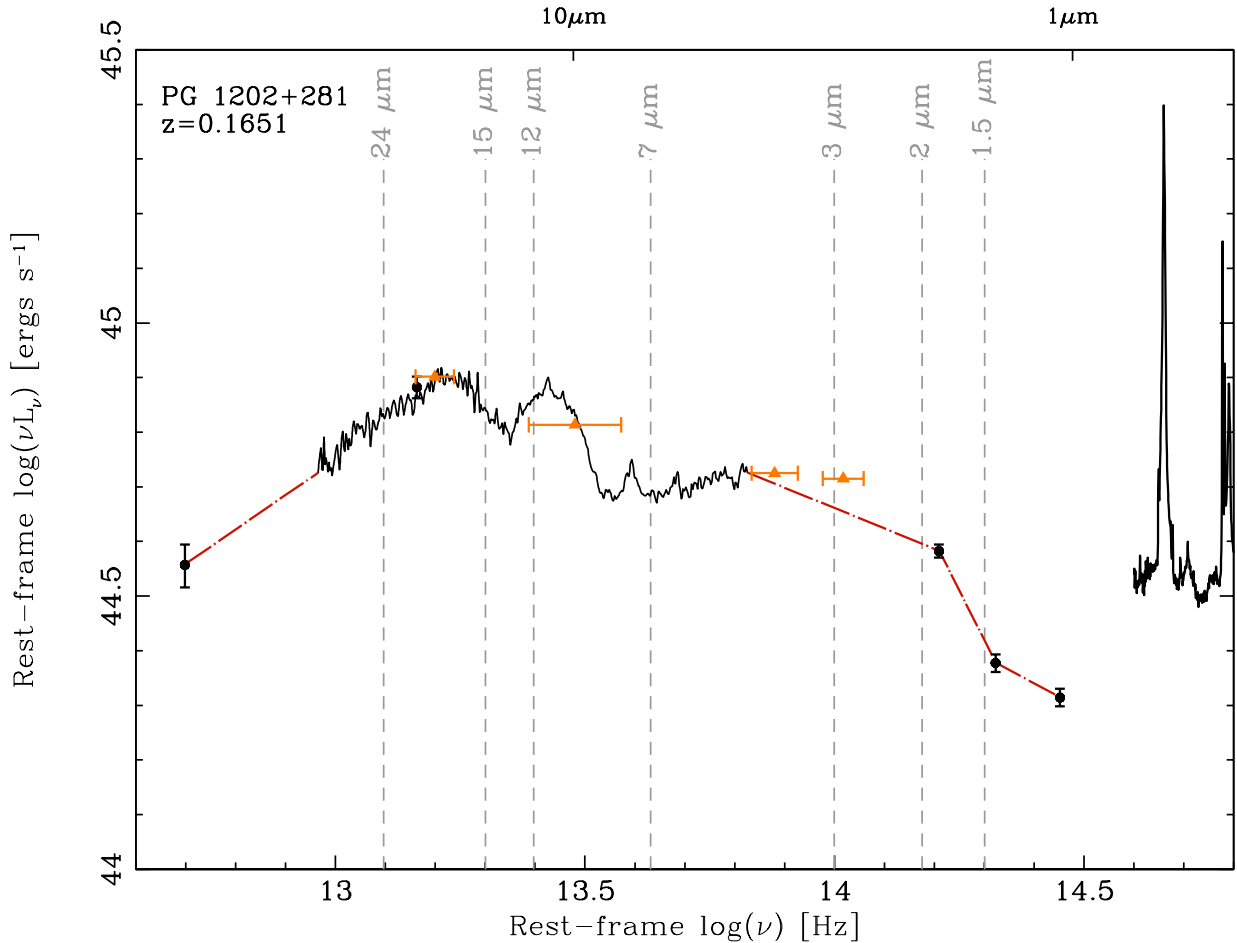}
\end{center}
\caption{The rest-frame SED of PG 1202+281.  The data are shown by black lines or black points for MIPS and 2MASS.  The red dotted-dashed line shows the interpolations used to bridge gaps in data.  The locations where monochromatic luminosities were measured are denoted by dashed gray lines.  WISE data points for this object are shown by solid gold triangles with horizontal bars indicating the frequency range of the WISE filter.  The W3 and W4 WISE fluxes have been corrected according to \citet{wright10}.}
\label{fig:SED}
\end{figure*}

We do not include WISE data in the SEDs.  For many objects we have IRS and MIPS coverage from {\it Spitzer} and do not gain new coverage by adding WISE, though we do find good agreement between the {\it Spitzer} and WISE data as evidenced by the low variability between the epochs of {\it Spitzer} and WISE and the WISE data plotted for one object in Fig.~\ref{fig:SED}.  WISE data are available for $\sim20$ objects with bolometric luminosities that do not have {\it Spitzer} data.  While adding these objects would increase the numbers in the sample at some wavelengths, it would also increase the uncertainty.  This would detract from the strength of these SEDs, namely the good coverage and detail, particularly of the emission features in the infrared that would be wiped out by interpolating between WISE points.

\subsection{Bolometric luminosities}
We calculated bolometric luminosities in \citet{runnoe12a}.  Of the total 85 objects in the atlas, only 63 have the appropriate wavelength coverage defining a smaller bolometric luminosity sample with 23 RQ and 40 RL quasars.  The RQ quasars are all from either the PGX or FUSE-HST samples and are lower redshift ($z<0.5$), while the RL quasars come primarily from the RLQ sample and more than half have higher redshifts ($z>0.5$).  The RL quasars have an average luminosity about 6 times higher than those that are RQ.  Both RL and RQ quasars span approximately 2 orders of magnitude in luminosity.

Bolometric luminosities were calculated by integrating the SEDs from 1 $\mu$m to 8 keV under the assumption of isotropy.  The 1 $\mu$m lower limit was chosen specifically to avoid double counting infrared photons under this assumption and the 8 keV upper limit was chosen perforce because the data do not extend past this limit.  Gaps in data coverage in the near-infrared and extreme-ultraviolet were covered by interpolating a log-log power-law spectrum between data points on either side.  The resulting bolometric luminosities have the range log$(L_{bol})= 45.1-47.3$ and are listed in Table~\ref{tab:sample}.  Uncertainty in the bolometric luminosity has the potential to be large, $30-40$\% on average, and is likely systematic as well.  Since the intrinsic SED in that region is unobserved the precise value of the uncertainty is unknown.

\subsection{Monochromatic infrared luminosities}
We measured infrared monochromatic luminosities at 1.5, 2, 3, 7, 12, 15, and 24 microns.  These wavelengths are chosen to enable comparisons between our bolometric corrections and those in the literature and to facilitate the use of our corrections for objects observed with WISE or {\it Spitzer} at a range of redshifts.

For measuring infrared monochromatic luminosities, we defined a smaller infrared sample of 37 objects that have data coverage from IRS and MIPS and bolometric luminosities from \citet{runnoe12a}.  In the infrared sample, 23 objects are RQ and 14 are RL and the RQ objects have a factor of 2 higher luminosity on average, though both RL and RQ quasars still span approximately 2 orders of magnitude in luminosity.  

\begin{figure}
\begin{center}
\includegraphics[width=8.9 truecm]{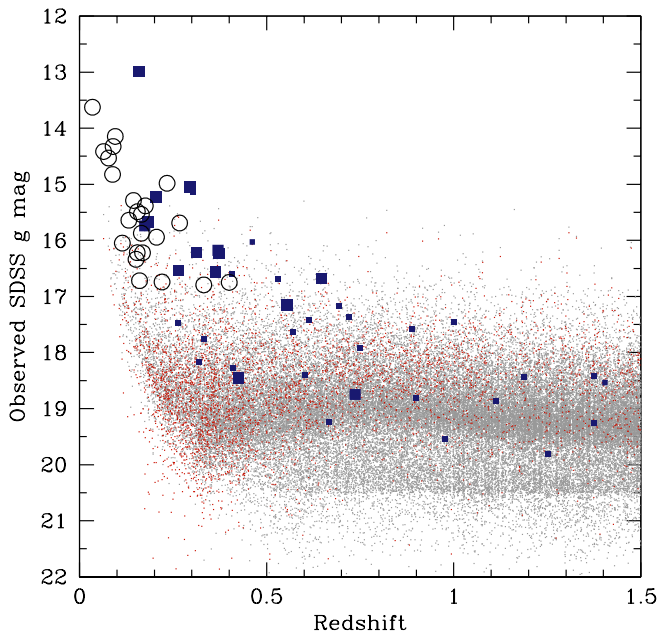}
\end{center}
\caption{The region of observed parameter space occupied by the 62 objects from the bolometric luminosity sample and the SDSS DR7 quasar catalog with the infrared sample distinguished.  SDSS quasars are small gray points and SDSS quasars with WISE and 2MASS coverage are small red points. RL quasars from this work are solid blue squares and RQ quasars are open black circles, with the infrared sample distinguished by larger symbols.  The bolometric corrections derived here may lose accuracy for objects that occupy another region of parameter space.}
\label{fig:sample}
\end{figure}

We compare our samples in observed parameter space to the Sloan Digital Sky Survey (SDSS) Data Release 7 (DR7) quasar catalog \citep{schneider10} in Fig.~\ref{fig:sample}.  We calculated approximate SDSS g magnitude for our objects by converting the flux at the center of the g-band to a magnitude with the SDSS flux zero point.  SDSS quasars that have coverage in WISE and 2MASS are highlighted.  In WISE, the bolometric luminosity sample covers the magnitude range $8.4-15.5$, $7.4-14.0$, $5.1-11.1$, and $2.9-8.8$ at 3.4, 4.6, 12, and 22 microns, respectively.  The bolometric corrections derived here will be most accurate for objects that lie in the region of parameter space occupied by the sample from which they were derived and may be less reliable when used on dissimilar objects.

There were gaps in data coverage in the infrared that had to be filled before monochromatic luminosities could be measured.  Gaps bewteen 2MASS data points, between the longest wavelength 2MASS point and the beginning of the IRS spectrum, and between the end of the IRS spectrum and the next MIPS photometry point were filled by interpolating a power-law spectrum between the existing data.  

These interpolations can be seen in Fig.~\ref{fig:SED} along with the locations selected for making bolometric corrections for one representative SED.

A log-log power-law spectrum may not perfectly describe the emission in the regions without data, so there is some uncertainty associated with this treatment of the SED.  We quantify the uncertainty in this step by computing monochromatic luminosities again using a WISE composite (S. Cales, private communication, 2012) scaled to the short-wavelength end of the IRS spectrum instead of a log-log power-law interpolation.  On average, the percent different between monochromatic luminosities at 3 $\mu$m measured using the WISE composite and the log-log power-law spectrum is 30\%, with the luminosities estimated from the WISE composite being smaller.  The WISE composite is visually indistinguishable from a log-log power-law spectrum in the region where we make our interpolation, so this uncertainty is likely an upper limit due to the fact that individual objects have different slopes than the WISE composite in the gap region.  For objects with slopes similar to the WISE composite in the gap region, the percent difference between monochromatic luminosities at 3 $\mu$m using the WISE composite and the log-log power-law spectrum is on the order of a few percent.

Contamination from the host galaxy can be a concern at infrared wavelengths.  \citet{shang11} corrected the 2MASS photometry for host contamination.  Without a prescription for separating the host and AGN emission at longer infrared wavelengths, we assumed that the SEDs were dominated by the AGN for the corrections at $3-24$ $\mu$m.  This assumption is likely a good one as the host component does not contribute significantly to the total emission until wavelengths longer than 24 $\mu$m \citep{netzer07,mullaney11}

The monochromatic luminosities, expressed as $\lambda L_{\lambda}$, were then measured from the SED.  The fluxes are well known and so uncertainties are small on these measurements.  Monochromatic luminosities at 1.5 $\mu$m are measured for 62 objects in the bolometric luminosity sample and at 2, 3, 7, 12, 15, and 24 $\mu$m for 37 objects.    

\input{sample.tex}

\section{Deriving infrared bolometric corrections}
\label{sec:analysis}
In this section we discuss the derivation of bolometric corrections cast in several mathematical forms.  We determined the ratio $\zeta_{ratio}=L_{iso}/\lambda L_{\lambda}$ for each object and fit lines to log$(L_{iso})$ versus log$(\lambda L_{\lambda})$ to determine bolometric corrections with zero intercepts, $L_{iso}=\zeta\, \lambda L_{\lambda}$, and with nonzero intercepts, log$(L_{iso})=A+B$ log$(\lambda L_{\lambda})$.  We make all three corrections at 1.5, 2, 3, 7, 12, 15, and 24 $\mu$m.  See \S \ref{sec:discussion} for a discussion about which bolometric corrections are the best to use.

\subsection{Ratio bolometric corrections}	
\begin{figure*}
\begin{minipage}[!b]{5.25cm}
\centering
\includegraphics[width=6cm]{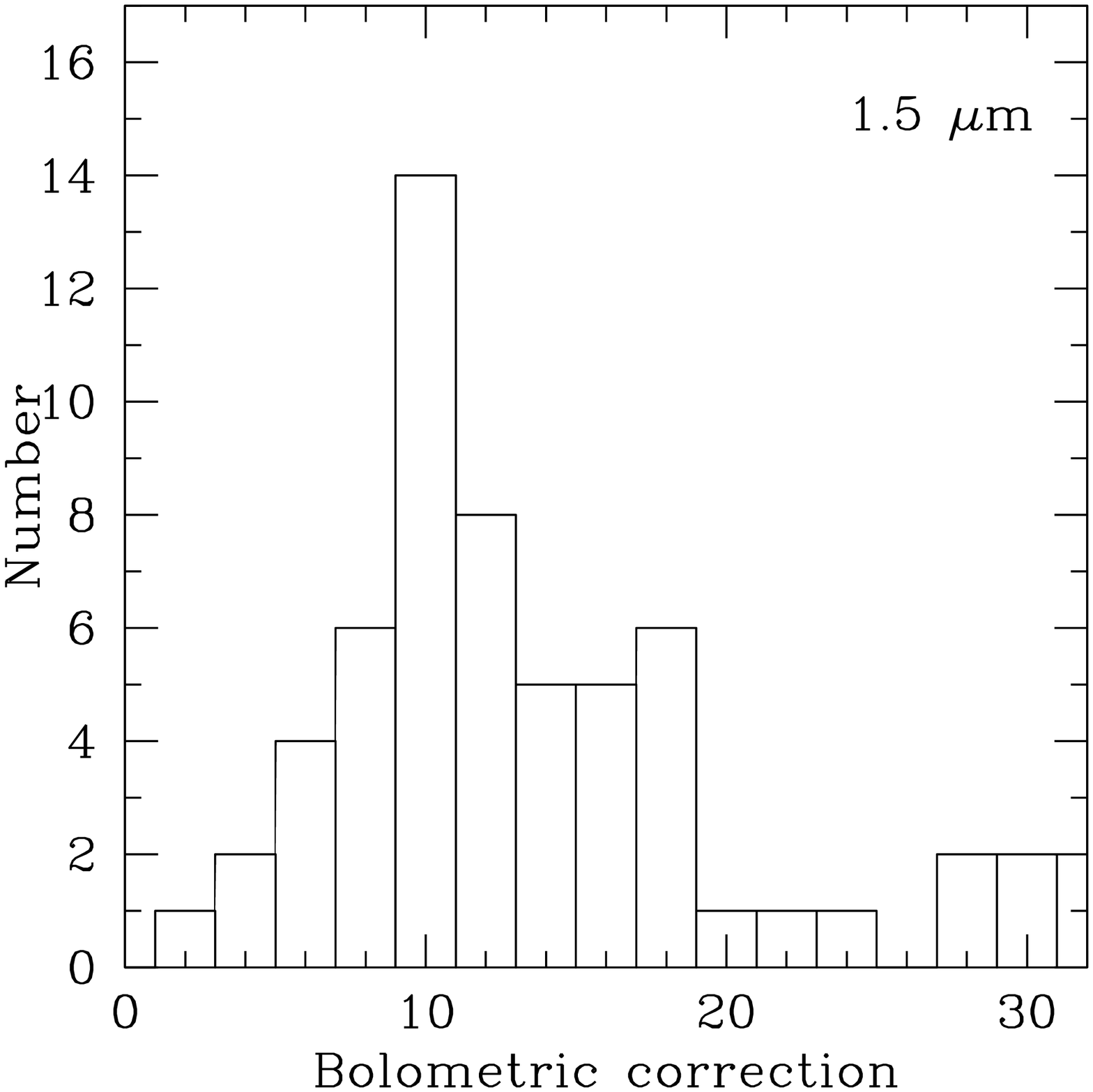}
\end{minipage}
\hspace{0.6cm}
\begin{minipage}[!b]{5.25cm}
\centering
\includegraphics[width=6cm]{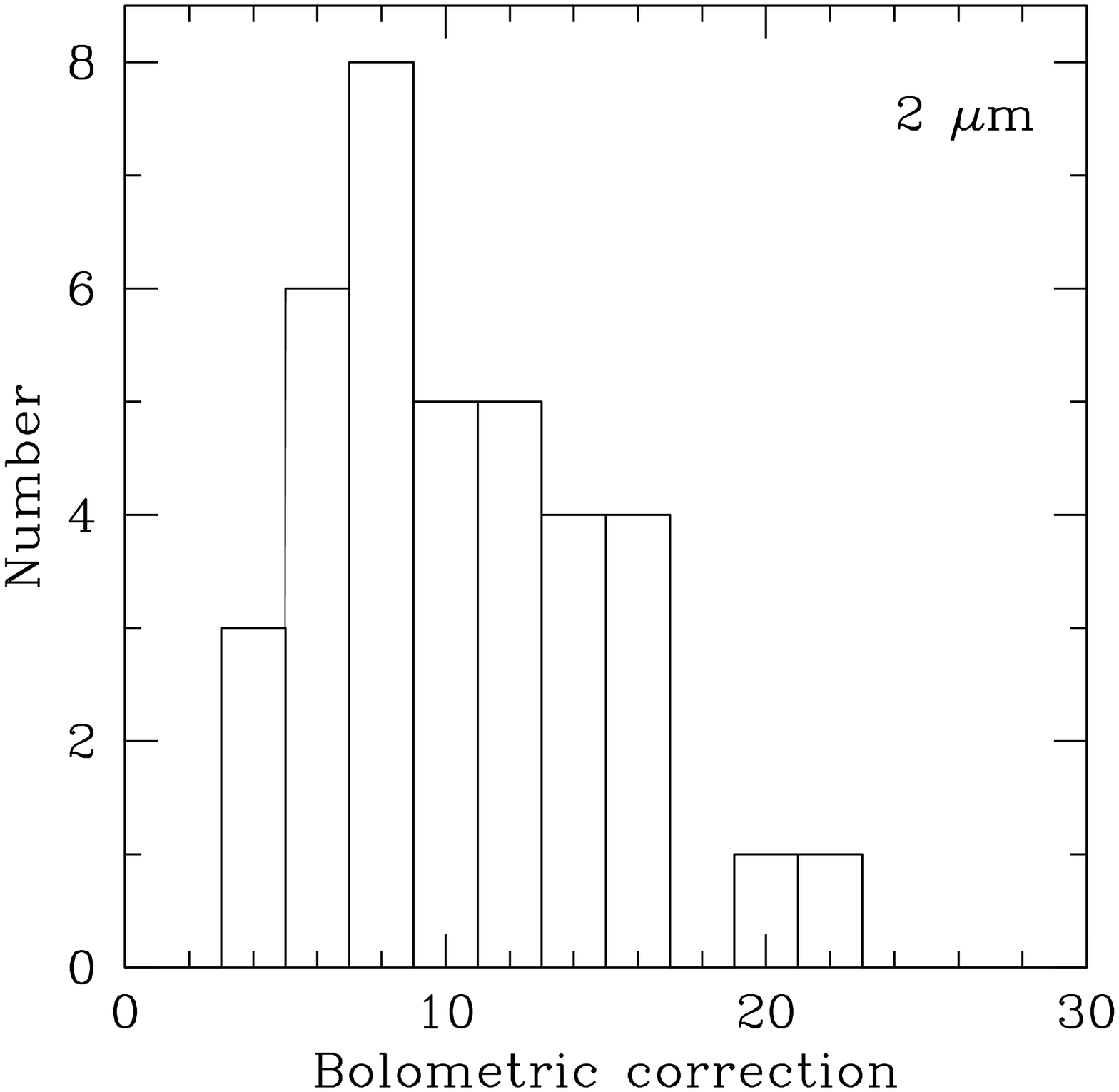}
\end{minipage}
\hspace{0.6cm}
\begin{minipage}[!b]{5.25cm}
\centering
\includegraphics[width=6cm]{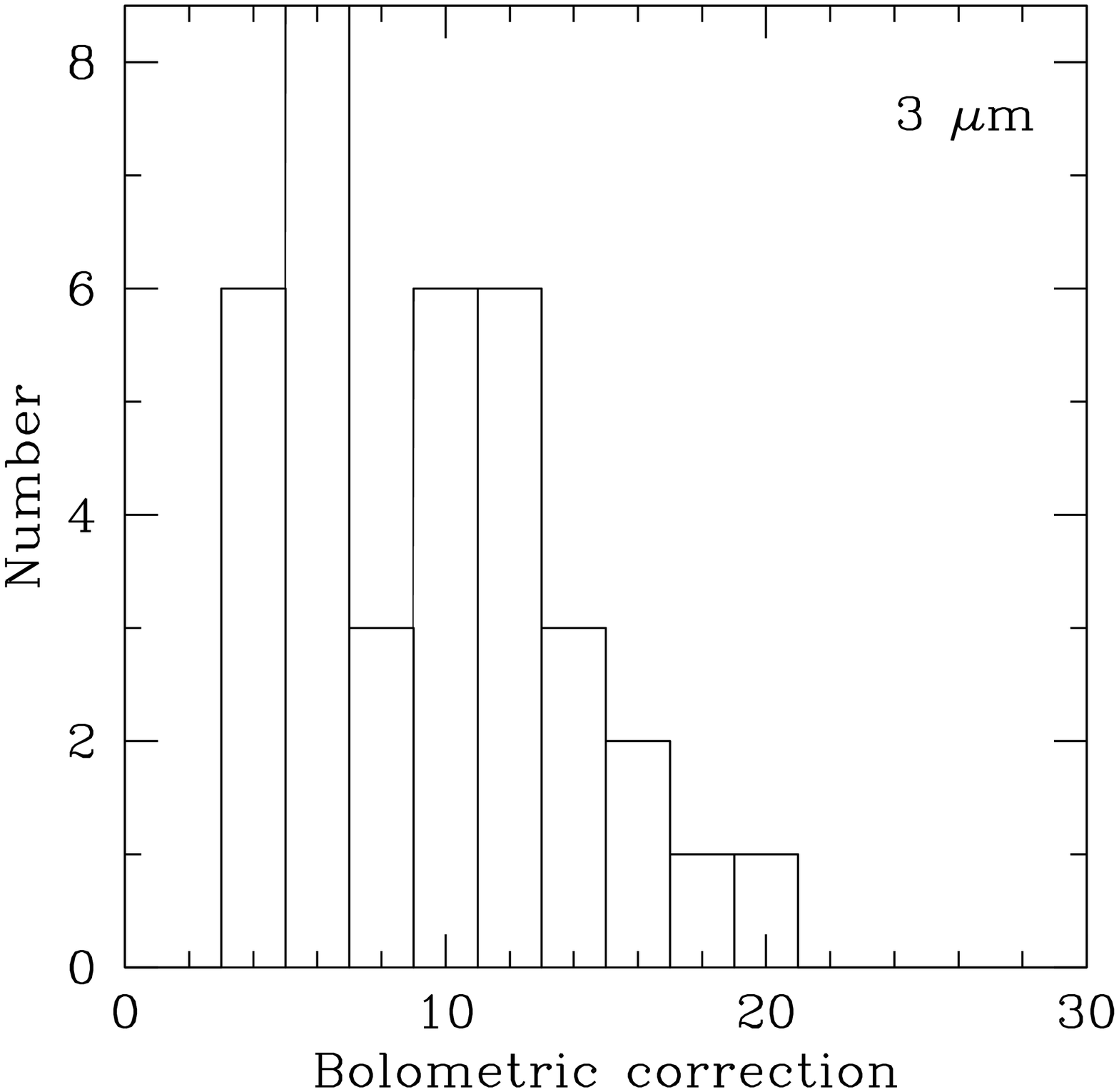}
\end{minipage}
\begin{minipage}[!b]{5.25cm}
\centering
\includegraphics[width=6cm]{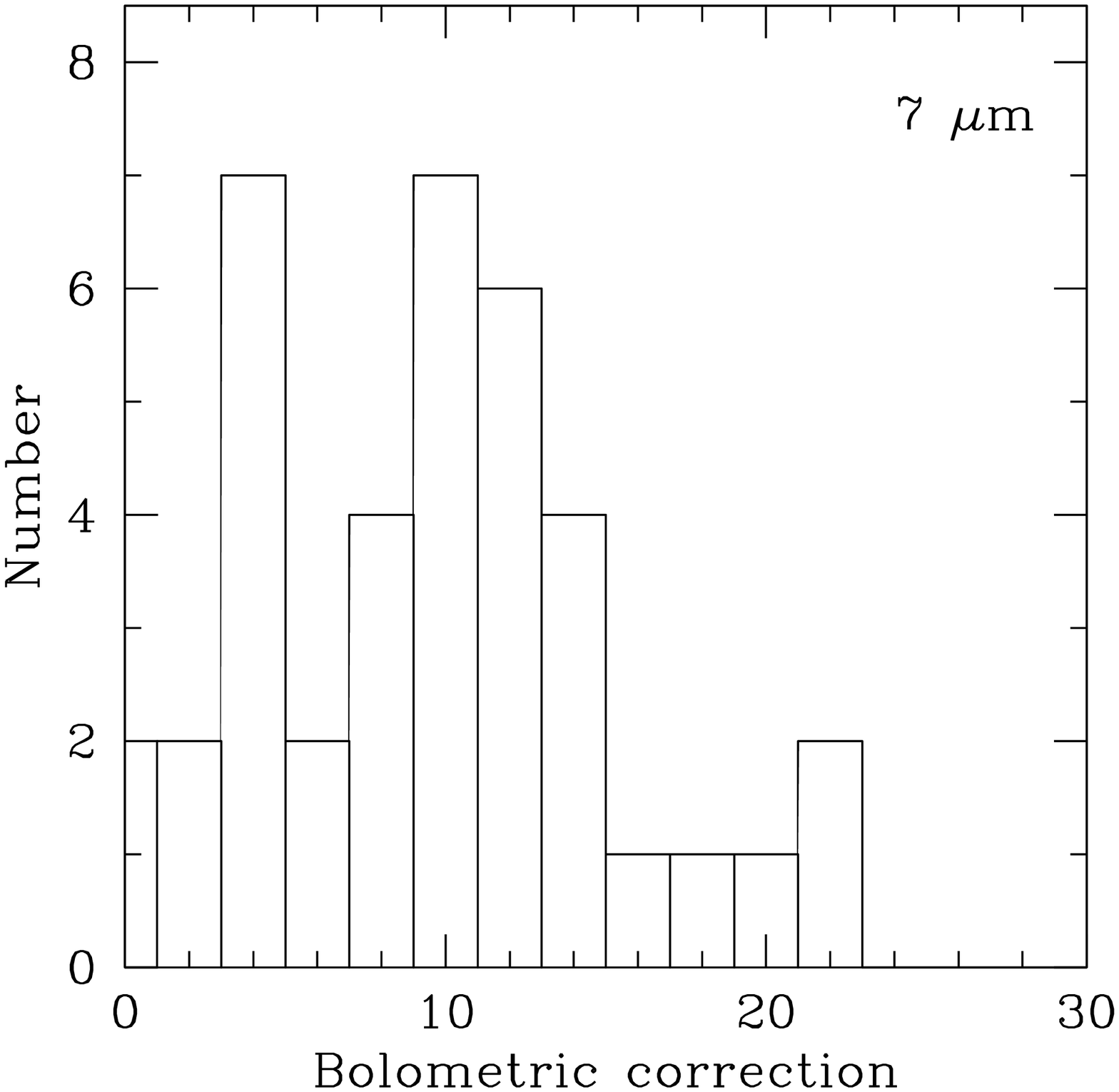}
\end{minipage}
\hspace{0.6cm}
\begin{minipage}[!b]{5.25cm}
\centering
\includegraphics[width=6cm]{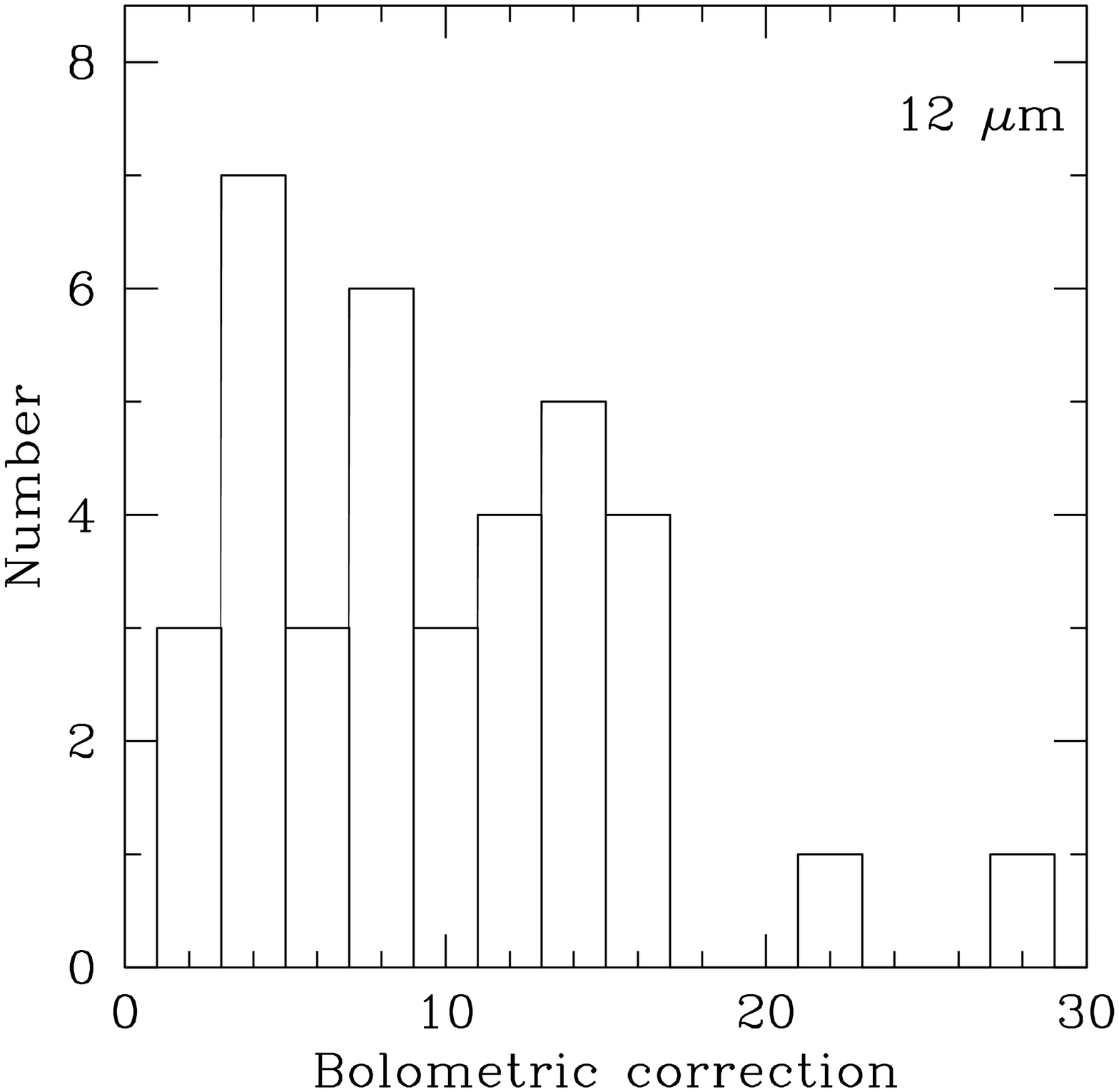}
\end{minipage}
\hspace{0.6cm}
\begin{minipage}[!b]{5.25cm}
\centering
\includegraphics[width=6cm]{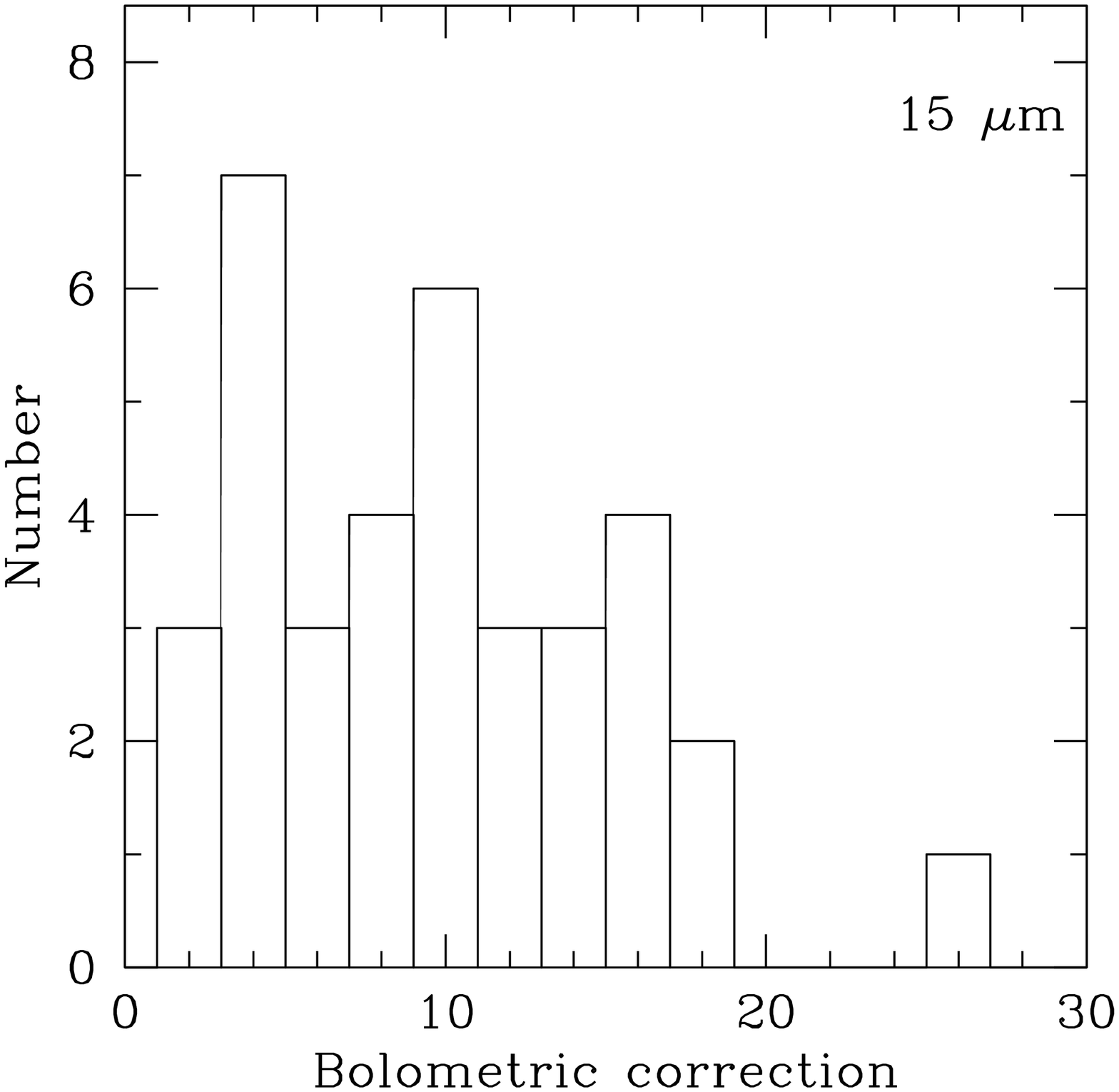}
\end{minipage}
\hspace{0.6cm}
\begin{minipage}[!b]{5.25cm}
\centering
\includegraphics[width=6cm]{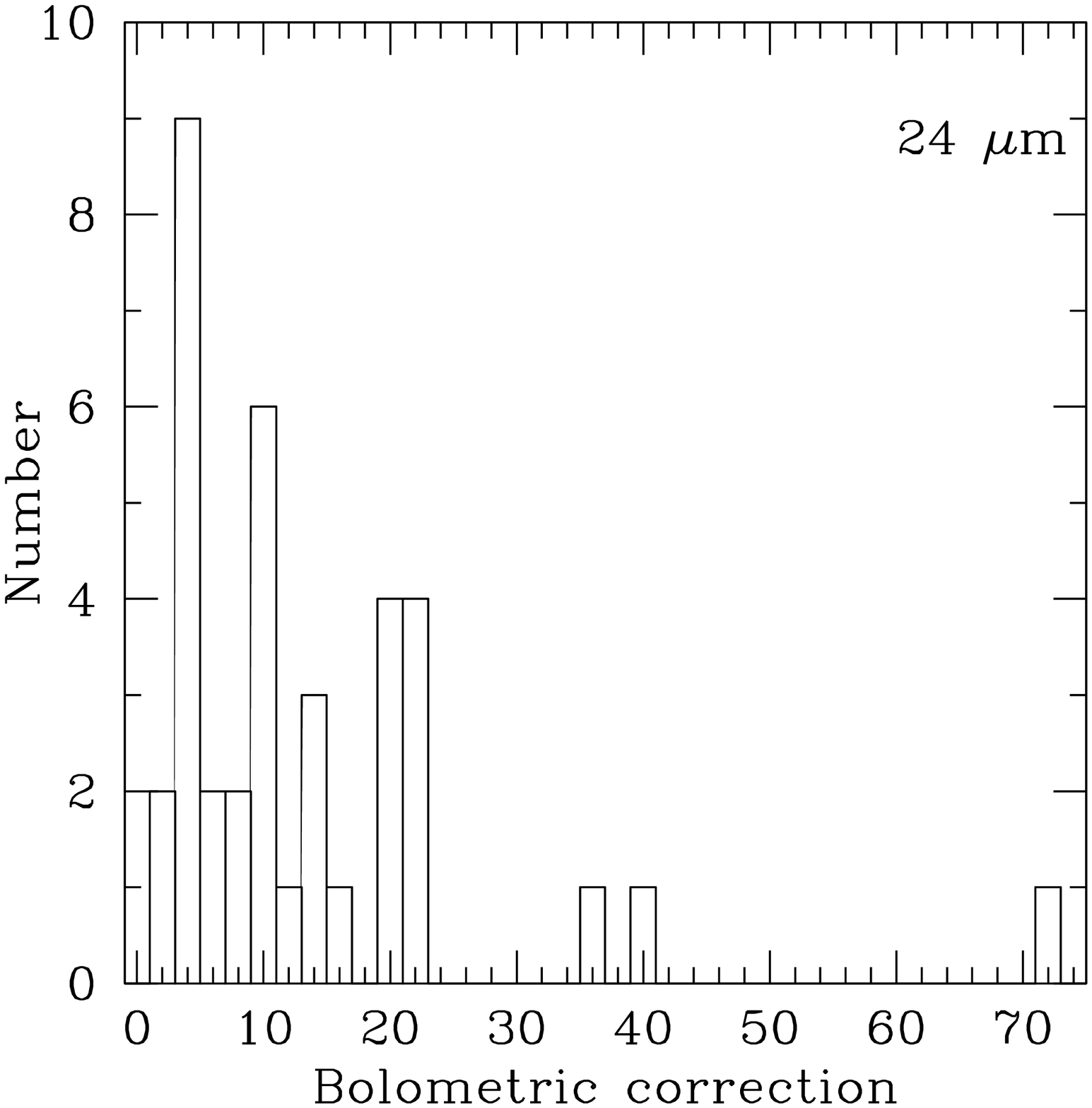}
\end{minipage}
\caption{Histograms of bolometric corrections at 1.5, 2, 3, 7, 12, 15, and 24 $\mu$m.  There are 62 objects total for corrections at 1.5 $\mu$m and 37 objects for corrections at all other wavelengths.  The histogram for 24 $\mu$m has much larger dispersion due to several object with low 24 $\mu$m luminosities.   \label{fig:bolhist}}
\end{figure*}

The ratio of bolometric to monochromatic luminosity, $\zeta_{ratio}=L_{iso}/\lambda L_{\lambda}$, is the most traditional form of the bolometric correction.  This ratio is listed in Table~\ref{tab:sample} for each object and the distributions are given in Fig.~\ref{fig:bolhist}.  

The dispersion of infrared bolometric corrections is greater than optical/UV bolometric corrections, where the bolometric corrections are $\zeta=4.2\pm0.1$ at 1450 \AA\ and $\zeta=8.1\pm0.4$ at 5100 \AA, but less than 2-10 keV X-ray bolometric corrections, where the corrections are $\zeta=23.31\pm3.91$ at 2 keV for RL objects and $\zeta=88.99\pm30.18$ at 2 keV for RQ objects.  The distributions are fairly similar between the infrared wavelengths, except for the few outliers at 24 $\mu$m.  The large 24 $\mu$m bolometric corrections tend to result from objects that have particularly low 24 $\mu$m luminosities.  Statistics for these distributions are summarized in Table~\ref{tab:stats}. 

\input{stats.tex}

We fit a line to log$(L_{iso})$ versus log$(\lambda L_{\lambda})$ in order to determine a bolometric correction with a zero intercept.  A bolometric correction of this form assumes a single SED shape scaled to match a monochromatic luminosity but, because quasar SEDs vary in shape, there is no reason to assume a direct linear relationship with a zero intercept.  For this reason, we also fit a log-log power law with a nonzero intercept to log$(L_{iso})$ versus log$(\lambda L_{\lambda})$.

Fitting in this paper was done by minimizing the chi-squared statistic using \textsc{mpfit} \citep{markwardt09} which employs the Levenberg-Marquardt least squares method.  The uncertainties in $\lambda \,L_{\lambda}$ are negligible compared to those in $L_{iso}$ because the spectra are high signal-to-noise and the monochromatic fluxes are well known whereas the bolometric luminosities likely have large uncertainties due to the interpolation over the unobserved extreme-ultraviolet region of the SED.

Adding a nonzero intercept to the fit is a marginal to significant improvement depending on the wavelength, with t-ratios given in Table~\ref{tab:fits}.  At 1.5 and 24 $\mu$m the improvement is significant, whereas at $3-15$ $\mu$m the improvement is more marginal and at 1.5 $\mu$m the nonzero intercept may not be needed.  

\begin{figure*}
\begin{minipage}[!b]{7.7cm}
\centering
\includegraphics[width=7.7cm]{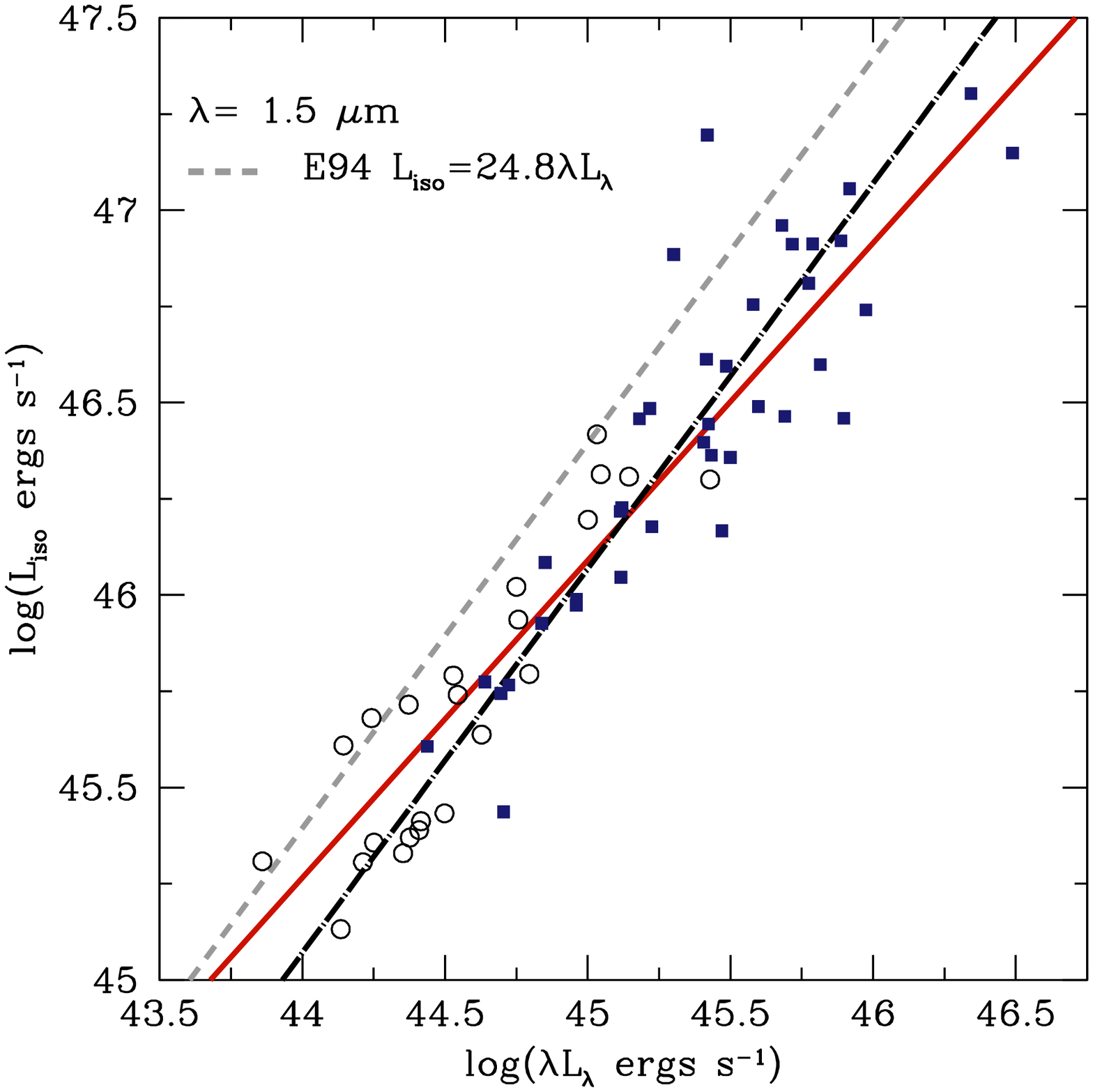}
\end{minipage}
\hspace{0.6cm}
\begin{minipage}[!b]{7.7cm}
\centering
\includegraphics[width=7.7cm]{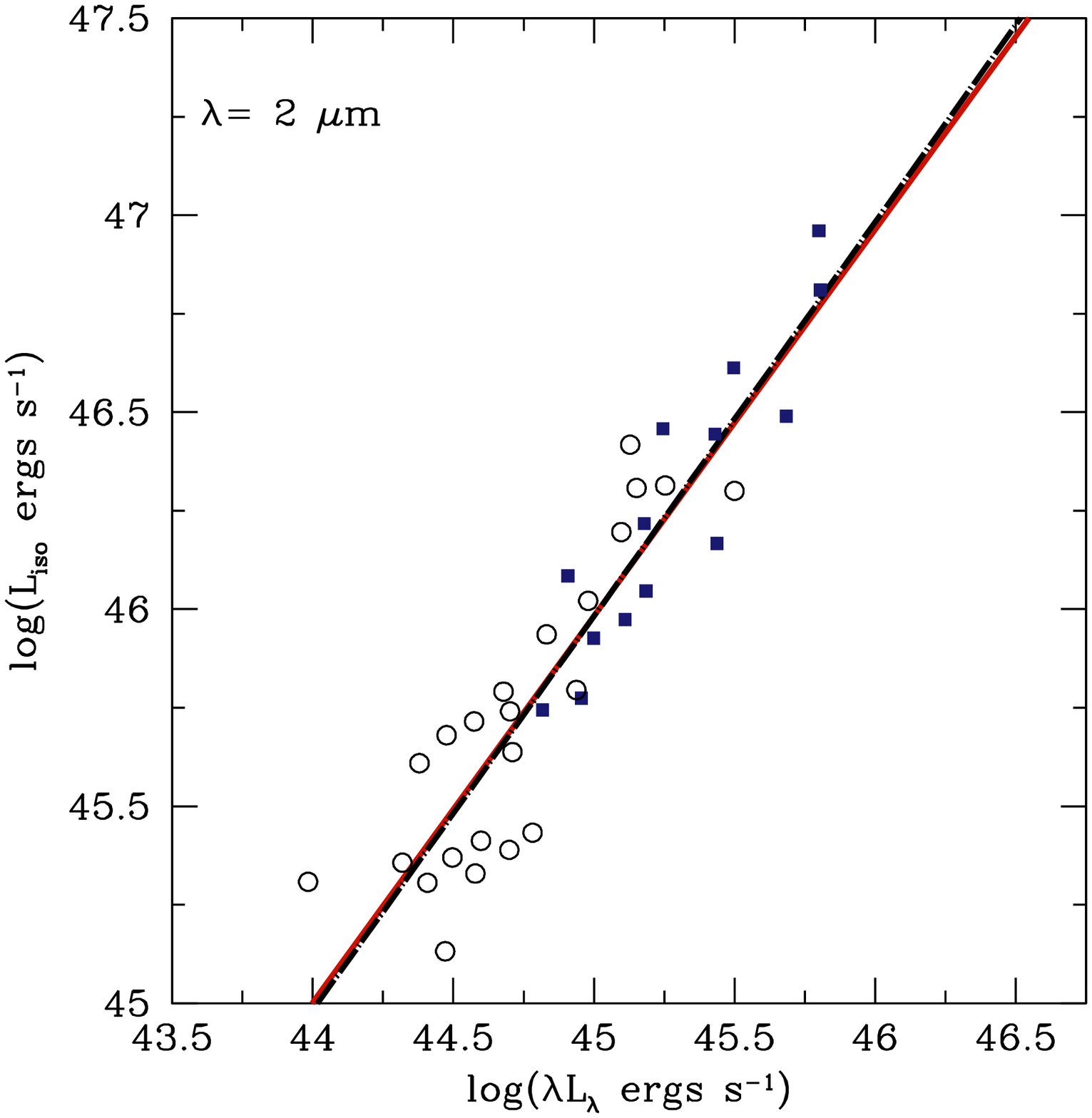}
\end{minipage}
\begin{minipage}[!b]{7.7cm}
\centering
\includegraphics[width=7.7cm]{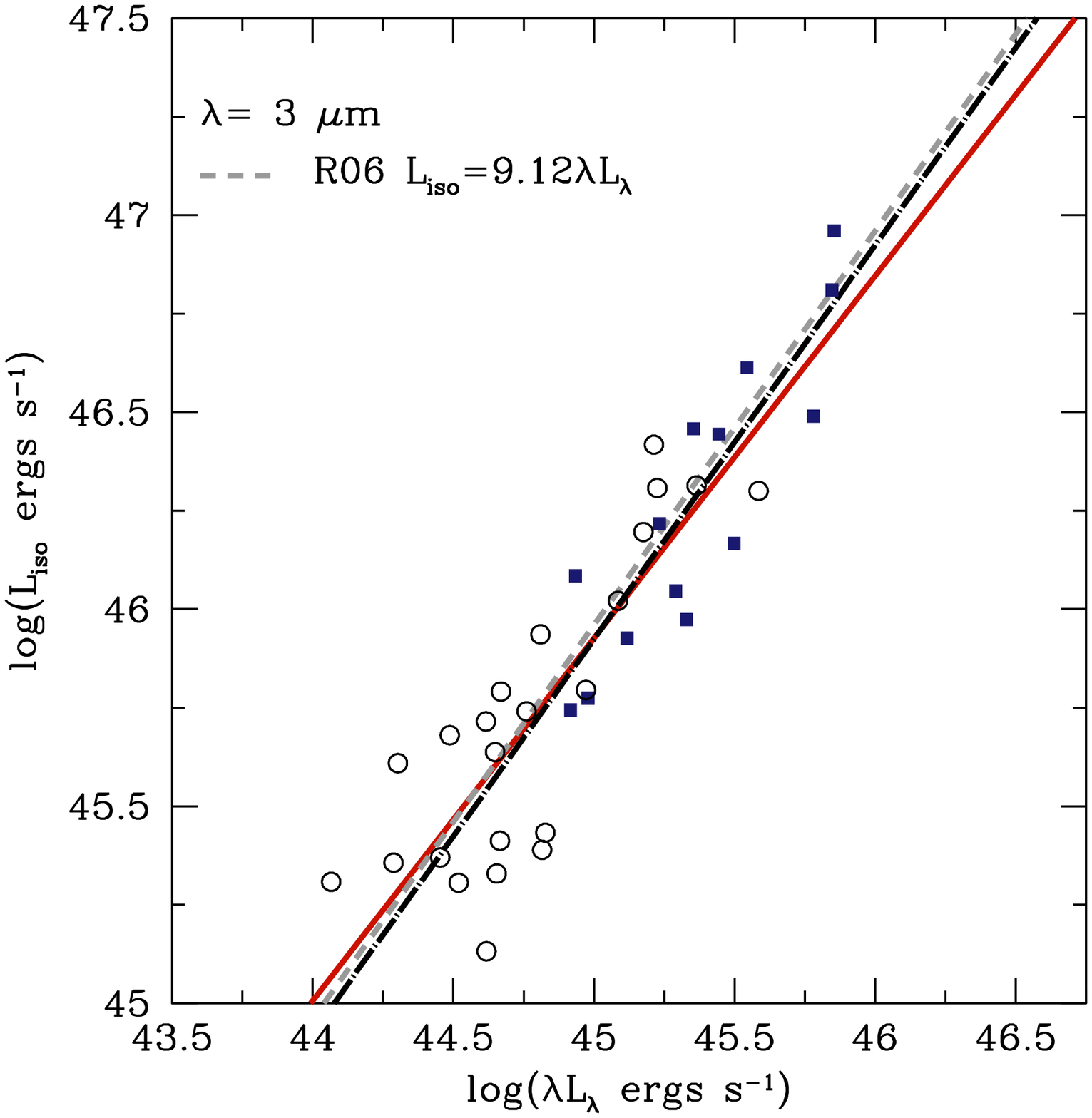}
\end{minipage}
\hspace{0.6cm}
\begin{minipage}[!b]{7.7cm}
\centering
\includegraphics[width=7.7cm]{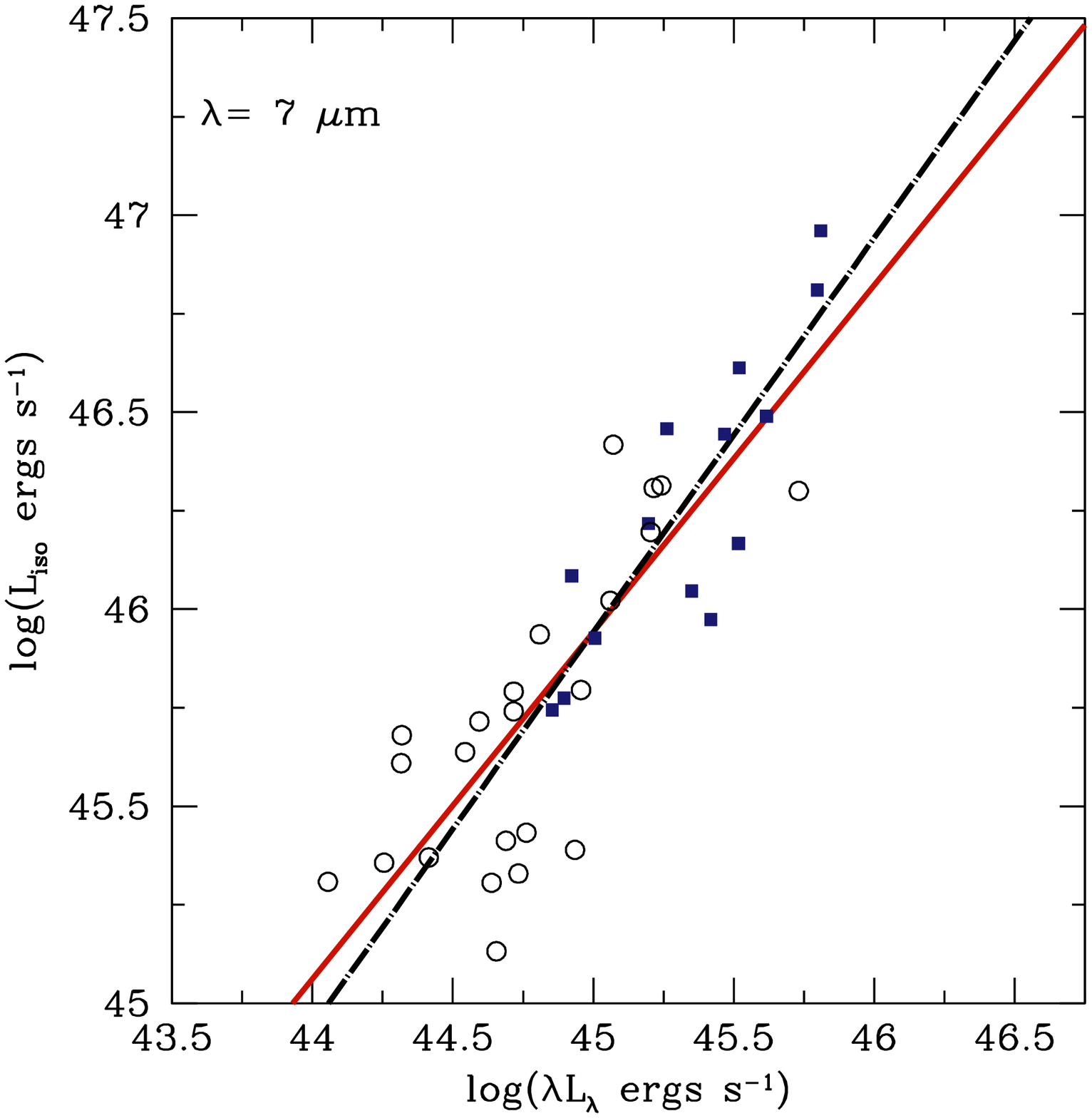}
\end{minipage}
\caption{log$(L_{bol})$ versus $\lambda$L$_{\lambda}$ with $\chi^2$ fits for 1.5, 2, 3, and 7 $\mu$m.  The dashed-dotted black lines indicate bolometric corrections with zero intercepts, and the solid red lines indicate bolometric corrections with nonzero intercepts. Filled blue squares indicate RL objects and open black circles indicate RQ objects.  62 objects were fit at 1.5 $\mu$m and 37 were fit at all other wavelengths.  The dashed gray lines are the \citet{elvis94} bolometric correction at 1.5 $\mu$m and the \citet{richards06} bolometric correction at 3 $\mu$m. \label{fig:bolfita}}
\end{figure*}

\begin{figure*}
\begin{minipage}[!b]{7.7cm}
\centering
\includegraphics[width=7.7cm]{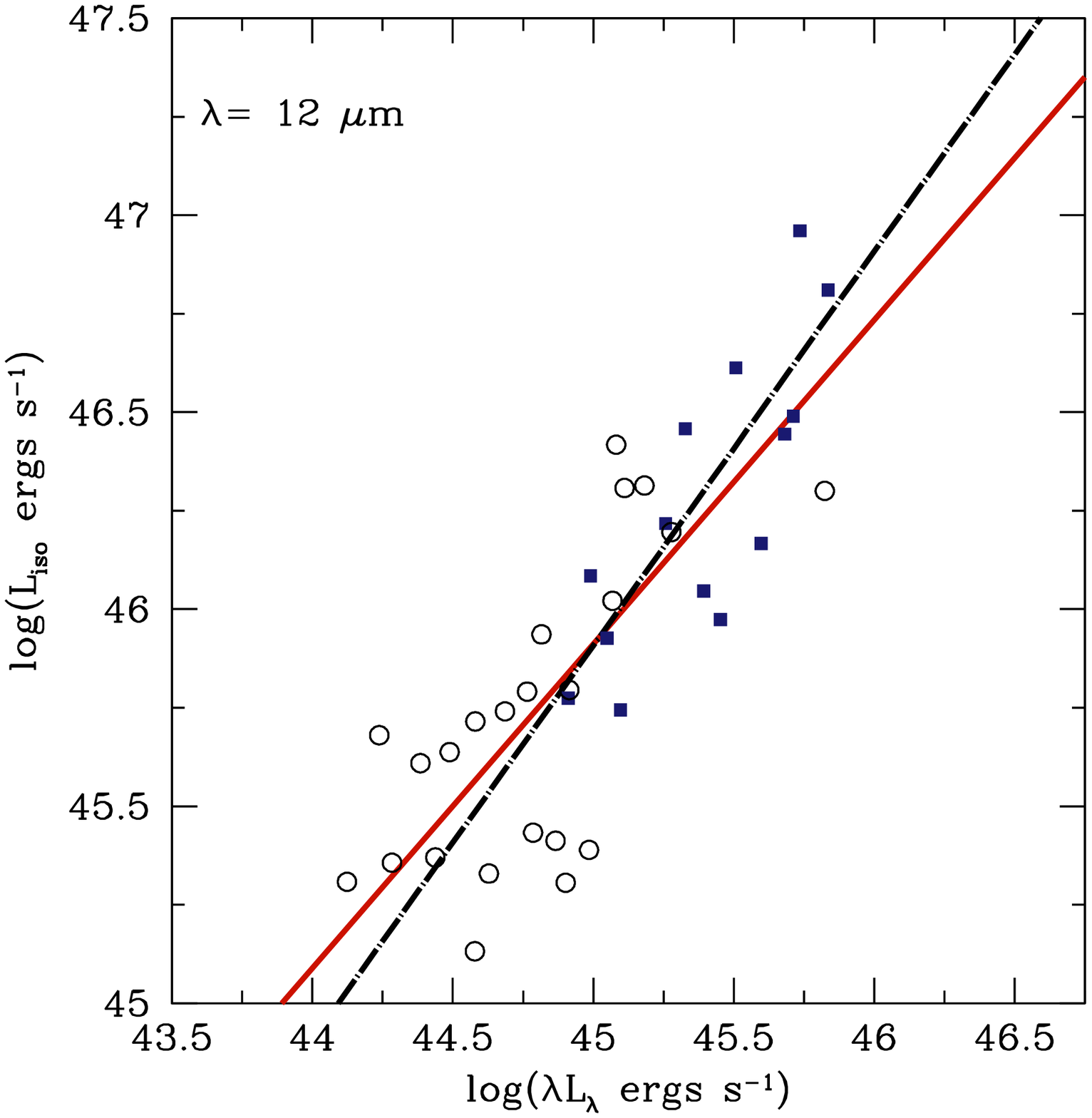}
\end{minipage}
\hspace{0.6cm}
\begin{minipage}[!b]{7.7cm}
\centering
\includegraphics[width=7.7cm]{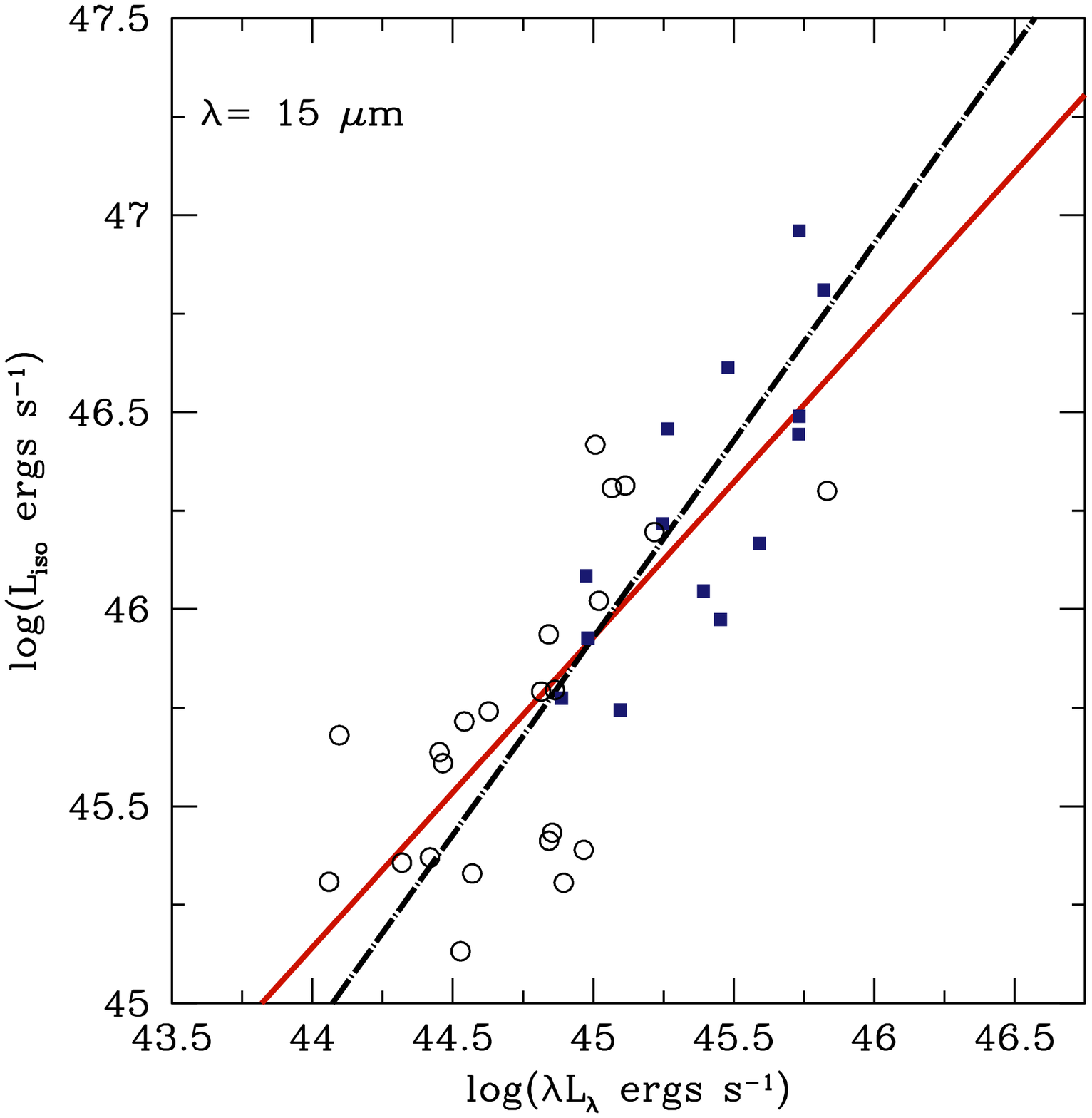}
\end{minipage}
\hspace{0.6cm}
\begin{minipage}[!b]{7.7cm}
\centering
\includegraphics[width=7.7cm]{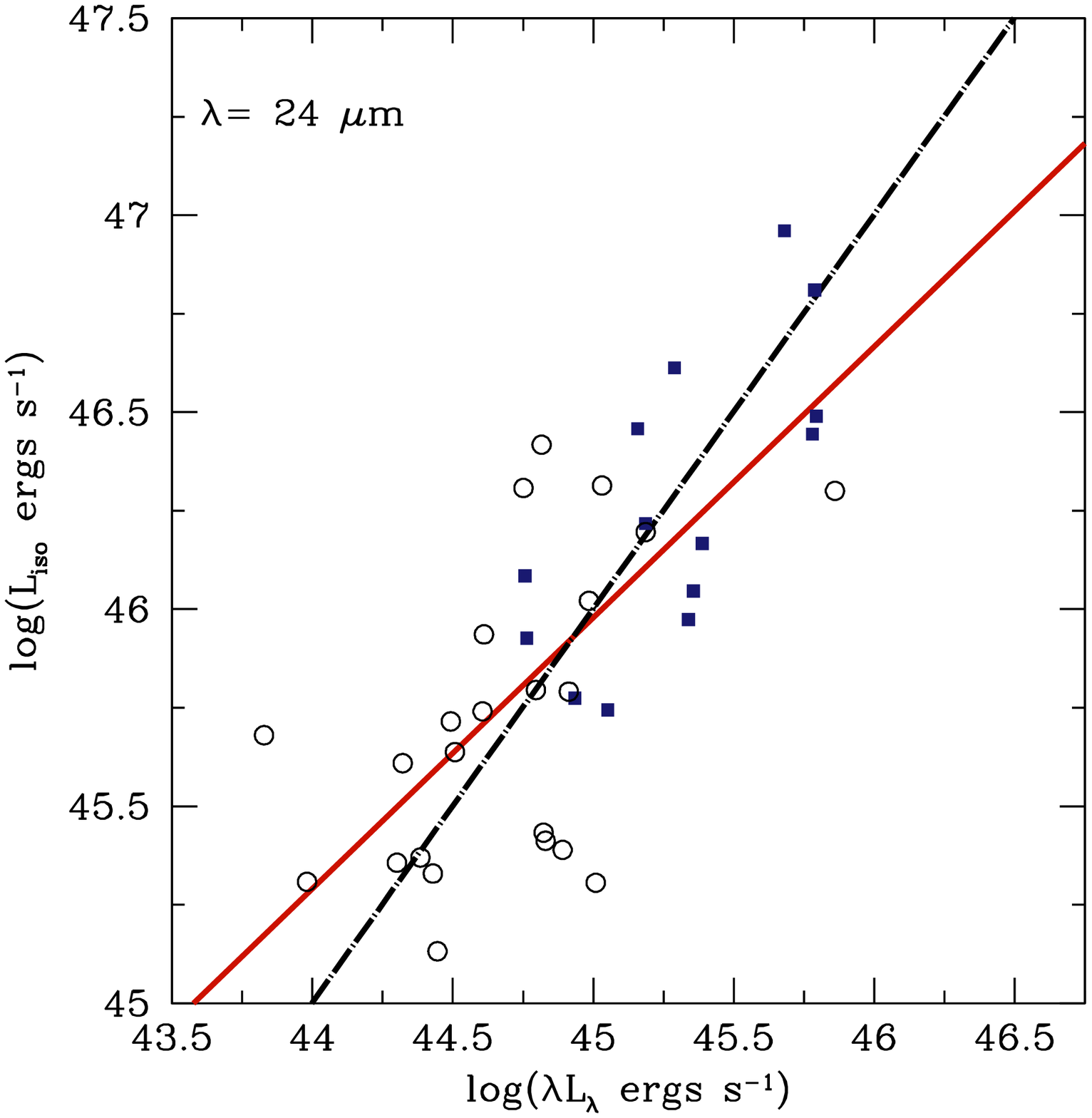}
\end{minipage}
\caption{log$(L_{bol})$ versus $\lambda$L$_{\lambda}$ with $\chi^2$ fits for 12, 15, and 24 $\mu$m.  The dashed-dotted black lines indicate bolometric corrections with zero intercepts, and the solid red lines indicate bolometric corrections with nonzero intercepts.  Filled blue squares indicate RL objects and open black circles indicate RQ objects.  37 objects were fit at these wavelengths. \label{fig:bolfitb}}
\end{figure*}

The fits are displayed in Figs.~\ref{fig:bolfita} and \ref{fig:bolfitb} and the bolometric corrections are given in Table~\ref{tab:fits}.  Uncertainties in the table are 1-sigma uncertainties in the fit coefficients.  


\input{corrections.tex}

\subsection{Radio class}
It is important to justify the use of a single bolometric correction for RL and RQ objects, which are known to have differences in their SEDs.  At infrared, optical, and UV wavelengths the SEDs of RL and RQ quasars are typically very similar, but at X-ray energies they distinguish themselves.  Because the X-rays are included in the bolometric luminosities, this may be a concern.

The differences between the X-ray emission of RL and RQ quasars are well known.  For a similar optical flux, RL quasars are typically three times more X-ray luminous \citep{zamorani81} than RQ quasars but can have an X-ray excess compared to RQ quasars as high as a factor of $3.4-10.7$ for extremely radio-loud quasars with R$^{*}>$ 3.5 \citep{miller11}.  

Despite these differences, in \citet{runnoe12a} we found that RL and RQ optical/UV bolometric corrections were consistent within the 95\% confidence intervals.

In order to investigate whether separate RL and RQ bolometric corrections are required in the infrared, we made fits of bolometric luminosity versus monochromatic luminosity to the RL and RQ objects separately.  Because the RL objects have higher luminosities on average for the bolometric luminosity sample, we isolated radio-loudness by fitting only the objects in the region where RL and RQ overlap in luminosity.  This is not an issue in the infrared sample.  

\begin{figure}
\begin{center}
\includegraphics[width=8.9 truecm]{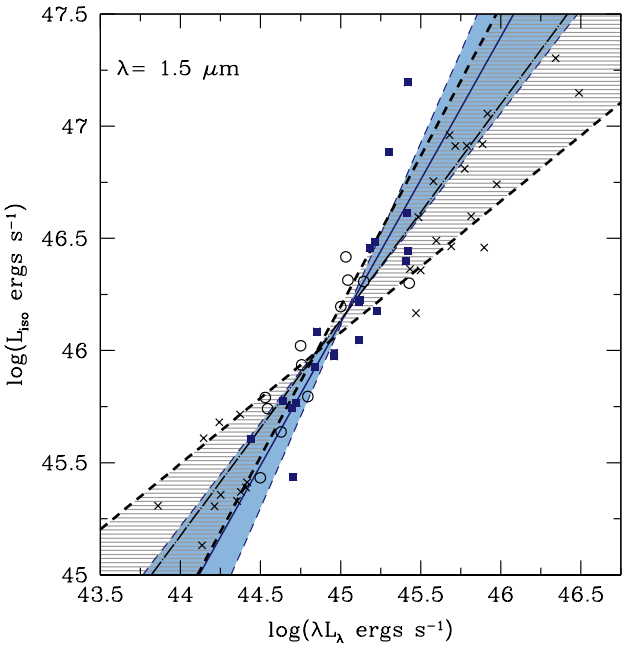}
\end{center}
\caption{Fits to bolometric versus monochromatic luminosity at 1.5 $\mu$m for RL and RQ objects of overlapping luminosity.  RL points in the region of overlapping luminosity are blue squares and RQ points in the region of overlapping luminosity are black circles.  Points outside of the region of overlapping luminosity are black crosses and are not included in the fits.  95\% confidence intervals are shaded in blue (solid) for the RL fit and in gray (hashed) for the RQ fit.  RL and RQ bolometric corrections are consistent within the 95\% confidence intervals at 1.5 $\mu$m.  The fit itself is a solid blue line for RL objects and a dashed-dotted black line for RQ objects.}
\label{fig:confidence}
\end{figure}

RL and RQ infrared bolometric corrections are consistent within the 95\% confidence intervals at all wavelengths considered in this investigation.  Fig.~\ref{fig:confidence} shows fits to bolometric versus monochromatic luminosity for RL and RQ objects of overlapping luminosity at 1.5 $\mu$m, which has the most restrictive confidence intervals due to the larger sample size.  We point out, because the slopes for RL and RQ objects do appear somewhat different in Fig.~\ref{fig:confidence}, that, while they are consistent within the 95\% confidence intervals, the confidence intervals are fairly large for such a small sample and a chance remains that separate corrections would be appropriate.  

Our result is consistent with the result from \citet{runnoe12a} that RL and RQ optical/UV bolometric corrections are consistent within the 95\% confidence intervals and we list here the same caveat.  The unknown emission in the extreme-ultraviolet and soft X-rays was given the same treatment in RL and RQ quasars, but if they have intrinsically different emission a different interpolation would be required.

The blazars originally included in this sample have been removed and these bolometric corrections are not appropriate for use on blazars.  Extreme caution should be used for radio-loud quasars expected of being blazars because of the higher likelihood of strong contamination from a jet-synchrotron component.  These bolometric corrections are likely not appropriate for such objects.

\section{Discussion}
\label{sec:discussion}

The infrared bolometric corrections in Table~\ref{tab:fits} yield bolometric luminosities that are consistent with bolometric luminosities derived from the optical/UV bolometric corrections of \citet{runnoe12a}.  The relationship between the bolometric luminosities estimated from IR and optical/UV corrections is consistent with a one-to-one relationship within the errors.  On average, the scatter between bolometric luminosities estimated from the infrared corrections with nonzero intercepts and bolometric luminosities estimated from the optical/UV corrections with nonzero intercepts is 0.21 dex.  The scatter for bolometric luminosity derived from bolometric corrections with zero intercepts is similar.  

\begin{figure*}
\begin{minipage}[!b]{8cm}
\centering
\includegraphics[width=8cm]{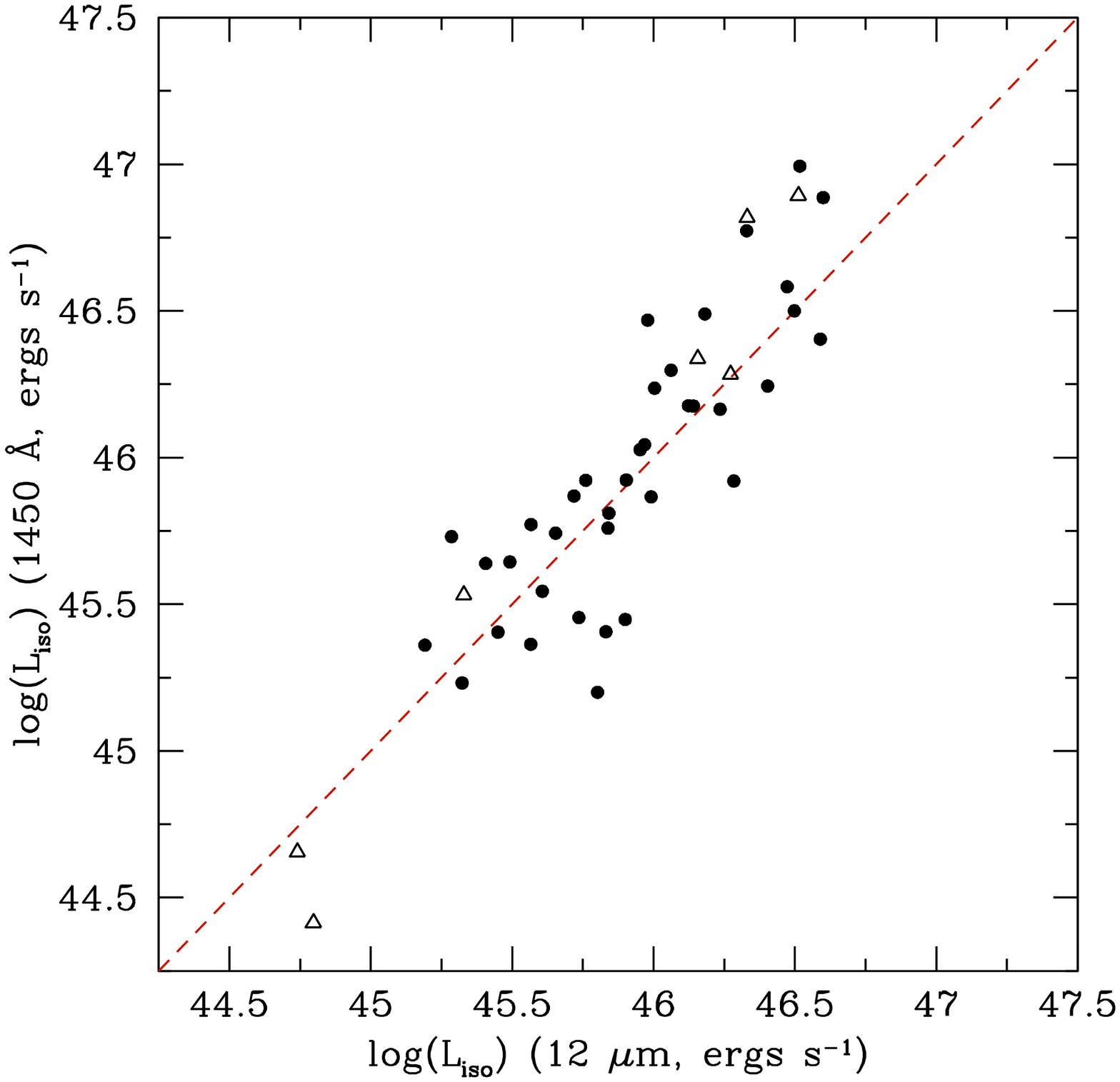}
\end{minipage}\hspace{0.6cm}
\begin{minipage}[!b]{8cm}
\centering
\includegraphics[width=8cm]{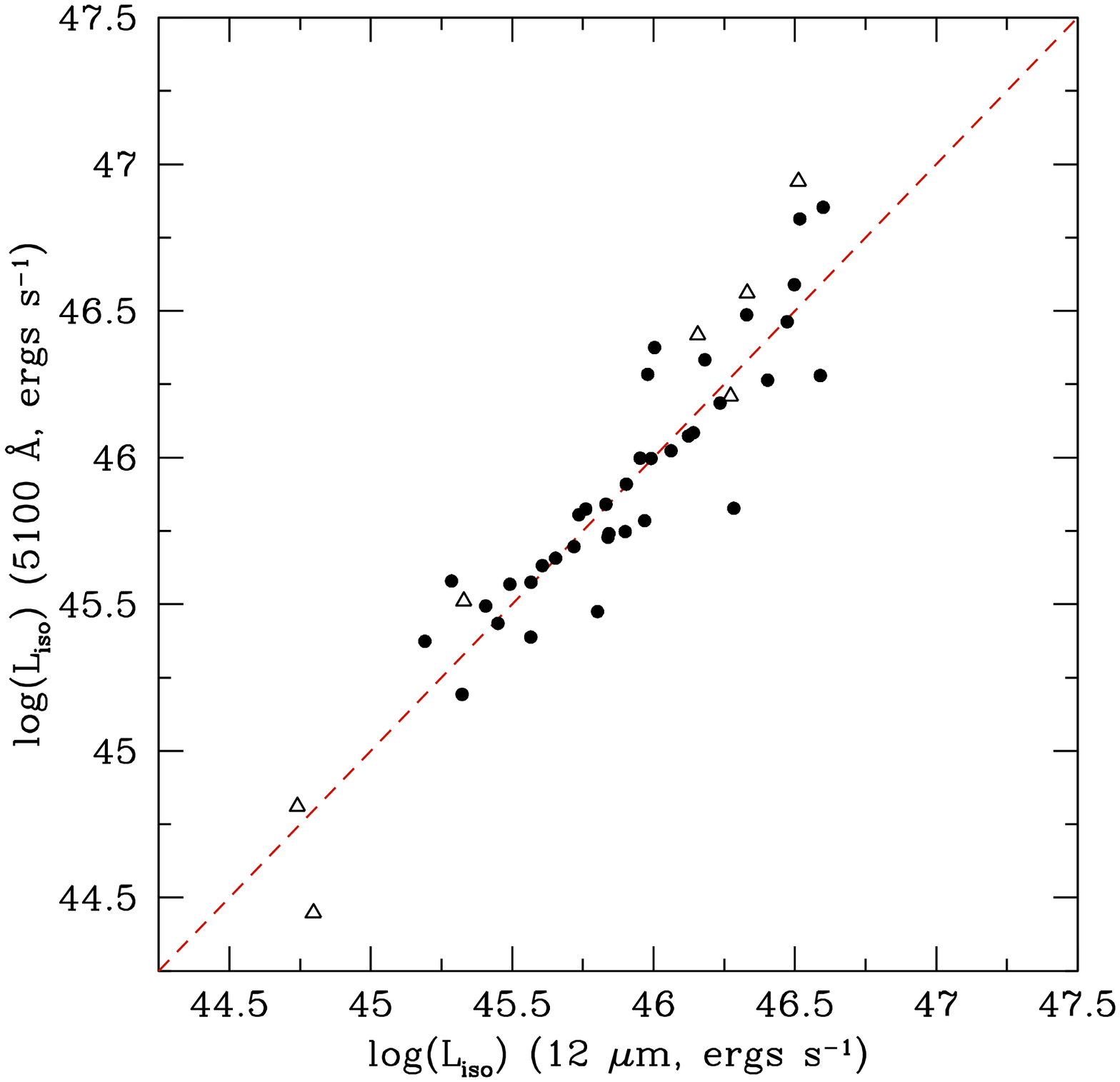}
\end{minipage}
\caption{Comparison of bolometric luminosities derived from optical/UV bolometric corrections of \citet{runnoe12a} and infrared bolometric corrections.  The left panel shows bolometric luminosities derived from the 1450 \AA\ correction with a nonzero intercept versus the 12 $\mu$m correction of the same type.  The right panel shows instead the 5100 \AA\ bolometric correction.  In both panels, the dashed red line shows where the bolometric luminosities are the same.  Open triangles show the good agreement between optical/UV and infrared bolometric corrections for objects not included in determining the corrections.  \label{fig:optIR}}
\end{figure*}

Fig.~\ref{fig:optIR} shows a comparison between optical/UV and infrared bolometric corrections.  Of the optical/UV bolometric corrections, the dispersion is minimized at 1450 \AA\ and maximized for 5100 \AA, so bolometric luminosities estimated from those corrections with nonzero intercepts are shown versus bolometric luminosities derived from the 12 $\mu$m correction with a nonzero intercept.  Comparisons between other wavelengths are similar.

There are 7 objects in the \citet{shang11} atlas that are not included in the bolometric luminosity sample because they lack X-ray data, but that do have optical/UV and infrared coverage.  These objects were not used to derive any bolometric corrections.  For these objects we measure monochromatic luminosities and apply the optical/UV and infrared bolometric corrections.  We find good agreement between bolometric luminosities estimated from optical/UV and infrared corrections in these objects.

The bolometric corrections derived here typically yield smaller bolometric luminosities than bolometric corrections in the literature.  For the specific corrections that we discuss, this is likely primarily the result of double counting the infrared when integrating the SED to determine bolometric luminosity.  Emission from the accretion disc is anisotropic, so for bolometric luminosity calculated under the assumption of isotropy, reprocessed infrared photons should not be included when integrating the SED.

\citet{elvis94} derives the median bolometric correction and standard deviation $L_{bol}=(24.8\pm8.4)\, \lambda L_{\lambda}$ at 1.5~$\mu$m.  This bolometric correction will overestimate bolometric luminosity by a factor of 2 compared to our 1.5 $\mu$m bolometric correction with a nonzero intercept for $\lambda L_{\lambda}=45.0$, which is in the middle of the range measured for our sample.  At the same monochromatic luminosity it will overestimate bolometric luminosity by a factor of 2.1 compared to our 1.5 $\mu$m bolometric correction with a zero intercept.  The \citet{elvis94} bolometric correction is not consistent with our 1.5 $\mu$m bolometric correction with a zero intercept within the given errors.

\citet{richards06} derives a bolometric correction and standard deviation $L_{bol}=(9.12\pm2.62)\, \lambda L_{\lambda}$ at 3 $\mu$m.  This bolometric correction will overestimate bolometric luminosity by a factor of 1.1 compared to our 3 $\mu$m bolometric correction with a nonzero or zero intercept for $\lambda L_{\lambda}=45.0$.  The \citet{richards06} bolometric correction is consistent with our 3 $\mu$m bolometric correction with a zero intercept within the given errors.

\begin{figure}
\begin{center}
\includegraphics[width=8.9 truecm]{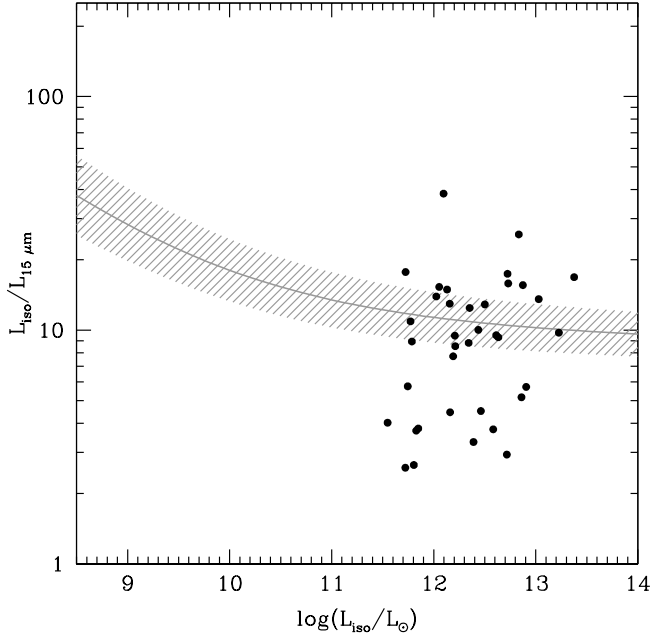}
\end{center}
\caption{The ratio $\zeta_{ratio}=L_{iso}/\lambda L_{\lambda}$ at 15 $\mu$m versus log$(L_{bol})$ for this sample with the luminosity dependent bolometric correction of \citet{hopkins07} with error shaded in gray.  The data are consistent with a luminosity dependent bolometric correction, but due to large dispersion and a small range in bolometric luminosity, one is not required by the data.}
\label{fig:hopkins}
\end{figure}

\citet{hopkins07} derives a luminosity dependent, double power law bolometric correction at 15 $\mu$m.  We measure $\lambda L_{\lambda}$ at 15 $\mu$m for our sample in order to make a direct comparison.  Fig.~\ref{fig:hopkins} shows the ratio $\zeta_{ratio}=L_{iso}/ \lambda L_{\lambda}$ at 15 $\mu$m versus log$(L_{bol})$ (similar to \citealt{hopkins07} fig. 1), with their double power-law bolometric correction over-plotted.  Our data appear consistent with a luminosity dependent bolometric correction, but the scatter is large enough and the range in $L_{bol}$ small enough that they by no means require it.

The dispersion in bolometric corrections is created by variation between individual SEDs, indicative of varying physical properties between objects.  Some of the spread in optical/UV bolometric corrections is due to viewing angle, this may also be true of infrared bolometric corrections.  The additional dispersion in infrared bolometric corrections beyond what is seen in the optical/UV may be due to variation in torus covering fraction.  If true, this indicates that there is a small range in torus covering fractions because there is only a small excess in dispersion of infrared bolometric corrections.

Because there are many bolometric corrections available in the literature, selecting the most appropriate one can be confusing.  Here we present our recommendations for selecting bolometric corrections.

\begin{itemize}
\item Bolometric corrections at different wavelengths are not all created equal; there is more dispersion around the mean correction at some wavelengths.  In some cases, the data do not allow a choice between bolometric corrections at different wavelengths, but when they do we recommend that the bolometric correction with the least dispersion be selected.  This means that UV corrections are preferred, then optical corrections, infrared corrections (with those at shorter wavelengths being preferred), and finally X-ray corrections.  However, in cases where optical/UV reddening can be identified it is better to use an infrared bolometric correction.

\item Bolometric corrections at a given wavelength come in different mathematical forms.  We recommend bolometric corrections with nonzero intercepts because regression analysis shows that adding a constant is usually significant.  We provide bolometric corrections with zero intercepts primarily for the purpose of comparing to the literature.

\item It is important to apply bolometric corrections to objects that are in the same region of parameter space for which the correction was derived.  Bolometric corrections from the bolometric luminosity sample of 62 objects are appropriate for objects with $z=0.03-1.4$ and log$(L_{bol})=45.1-47.3$.  Bolometric corrections from the infrared sample of 37 objects are appropriate for $z=0.03-0.74$ and log$(L_{bol})=45.1-47.0$.  For higher redshift or luminosity objects, it may be more appropriate to use the bolometric corrections of \citet{richards06}.

\item We recommend the additional correction to bolometric luminosity for the assumption of isotropy.  For a randomly selected sample viewed at the angle of $31^{\circ}$ suggested by \citet{barthel89}, bolometric luminosity will be over-estimated by about 33 per cent \citep[][fig. 11.]{nb10}  In \citet{runnoe12a}, we suggested the following correction, which should be applied after using infrared, optical/UV, or X-ray bolometric corrections:

\begin{equation}	
	L_{bol}=f\,L_{iso} \approx 0.75\, L_{iso}
\end{equation}

\noindent where $f$ is a factor describing the bias of an average viewing angle due to the anisotropic emission from a disc.  For objects to which the line of sight is known, a theoretical correction as a function of viewing angle can be found from \citet{nb10}, fig. 11.  Their predicted range for this correction to L$_{iso}$ is approximately 67 to 200 per cent for a relativistic disc including projection effects and limb darkening at viewing angles of zero to ninety degrees.  

Infrared emission is likely to be more isotropic than accretion disc emission, but it is still necessary to make this correction when using infrared bolometric corrections because they were derived from bolometric luminosities that included the isotropy assumption.

\item It will usually be necessary to make k-corrections before applying bolometric corrections in the infrared.  Because of the emission features in the infrared, we recommend scaling the appropriate composite SED from \citet{shang11} for filling large gaps in data rather than interpolating a log-log power-law spectrum between available data points.  Keep in mind when doing this, that the composite SEDs have had corrections for host galaxy emission applied to the 2MASS points in the near-infrared and to the optical spectrum.  This will change the shape of the SEDs at these wavelengths compared to uncorrected SEDs.  
\end{itemize}  

\section{Summary}
\label{sec:summary}
We have added infrared bolometric corrections to the suite of bolometric corrections derived from the \citet{shang11} SEDs.  We provide corrections with a zero and nonzero intercept for 1.5, 2, 3, 7, 12, 15, and 24 $\mu$m.  We find that corrections derived for RL and RQ objects are consistent within the 95\% confidence intervals and, while the data may be consistent with a luminosity dependent correction, they do not require one.

The bolometric corrections derived here are self consistent with the optical/UV bolometric corrections of \citet{runnoe12a}.  The scatter in infrared bolometric corrections is larger than in optical/UV bolometric corrections, but smaller than in X-ray bolometric corrections.  Bolometric luminosities resulting from these infrared corrections will typically be smaller than those resulting from those in the literature \citep[e.g.,][]{elvis94,richards06}, although the \citet{richards06} correction is consistent within the errors.

For objects with properties that fall in the region of parameter space covered by this sample (log(L$_{bol})=45.1-47.0$ for $2-24$ $\mu$m and log(L$_{bol})=45.1-47.3$ for 1.5 $\mu$m), we recommend the following infrared bolometric corrections:  
\begin{itemize}
\item {\bf Recommended Correction at 1.5 $\mu$m}
\input{fit1.tex}

\item {\bf Recommended Correction at 2 $\mu$m}
\input{fit2.tex}

\item {\bf Recommended Correction at 3 $\mu$m}
\input{fit3.tex}

\item {\bf Recommended Correction at 7 $\mu$m}
\input{fit7.tex}

\item {\bf Recommended Correction at 12 $\mu$m}
\input{fit12.tex}

\item {\bf Recommended Correction at 15 $\mu$m}
\input{fit15.tex}

\item {\bf Recommended Correction at 24 $\mu$m}
\input{fit24.tex}
\end{itemize}

\section*{Acknowledgments}
J. Runnoe would like to thank Sabrina Cales for providing the WISE composite spectral energy distributions.

This work funded by Wyoming NASA Space Grant Consortium, NASA Grant \#NNX10A095H.

\bibliographystyle{/Users/jrunnoe/Library/texmf/bibtex/bst/mn2e}
\bibliography{all.062712}
\clearpage

\label{lastpage}
\end{document}

%% file: commands.tex
\def\lsim{\lower0.3em\hbox{$\,\buildrel <\over\sim\,$}}
\def\gsim{\lower0.3em\hbox{$\,\buildrel >\over\sim\,$}}

%% file: irvar.tex
\begin{table*}
\begin{minipage}[2cm]{12cm}
\caption{Comparison between WISE and {\it Spitzer} \label{tab:spitzerwise}}
\renewcommand{\thefootnote}{\alph{footnote}}
\begin{tabular}{lccccrr}
Object & W3$_{{\it Spitzer}}$\footnotemark[1] & W4$_{{\it Spitzer}}$ & W3$_{\textrm{All-Sky}}$ &  W4$_{\textrm{All-Sky}}$ & $\Delta$W3\footnotemark[2] & $\Delta$W4 \\
       & (mag) 		  & (mag)  		      & (mag)	& (mag) 		  & (mag)  		      & (mag)	    \\
\hline
3C 263    &     8.28 &     6.21 &     8.26 &     6.12 &    -0.02 &    -0.09\\
3C 334    &     8.53 &     6.05 &     8.54 &     5.99 &     0.02 &    -0.06\\
3C 47    &     8.11 &     6.01 &     8.10 &     6.01 &    -0.00 &     0.00\\
4C 11.69   \footnotemark[3] &     7.48 &     5.32 &     8.71 &     6.59 & {\bf    1.23} & {\bf    1.27}\\
4C 34.47    &     7.48 &     5.38 &     7.53 &     5.41 &     0.05 &     0.03\\
IRAS F07546+3928    &     6.08 &     3.86 &     6.16 &     3.94 &     0.08 &     0.08\\
PG 0052+251    &     7.15 &     5.10 &     7.15 &     5.11 &     0.00 &     0.00\\
PG 0844+349    &     6.95 &     4.94 &     6.87 &     4.85 &    -0.07 &    -0.09\\
PG 0947+396    &     7.55 &     5.66 &     7.52 &     5.57 &    -0.02 &    -0.10\\
PG 0953+414    &     7.56 &     5.69 &     7.38 &     5.65 &    -0.18 &    -0.04\\
PG 1001+054    &     7.81 &     5.97 &     7.65 &     5.94 &    -0.15 &    -0.03\\
PG 1100+772    &     7.61 &     5.69 &     7.66 &     5.61 &     0.06 &    -0.08\\
PG 1114+445    &     6.63 &     4.53 &     6.64 &     4.48 &     0.01 &    -0.05\\
PG 1115+407    &     7.71 &     5.80 &     7.63 &     5.69 &    -0.07 &    -0.10\\
PG 1116+215    &     6.55 &     4.69 &     6.48 &     4.73 &    -0.07 &     0.05\\
PG 1202+281    &     7.29 &     5.11 &     7.44 &     5.07 &     0.15 &    -0.03\\
PG 1216+069    &     8.13 &     6.40 &     8.07 &     6.45 &    -0.07 &     0.04\\
PG 1226+023    &     4.88 &     2.86 &     4.98 &     3.05 &     0.10 &     0.19\\
PG 1309+355    &     6.96 &     4.69 &     7.02 &     4.74 &     0.05 &     0.05\\
PG 1322+659    &     7.70 &     5.73 &     7.71 &     5.71 &     0.01 &    -0.02\\
PG 1351+640    &     5.63 &     3.27 &     5.65 &     3.26 &     0.02 &    -0.01\\
PG 1352+183    &     8.11 &     6.00 &     7.95 &     6.05 &    -0.16 &     0.05\\
PG 1402+261    &     6.68 &     4.77 &     6.65 &     4.76 &    -0.03 &    -0.02\\
PG 1411+442    &     6.36 &     4.62 &     6.24 &     4.59 &    -0.12 &    -0.03\\
PG 1415+451    &     7.41 &     5.41 &     7.43 &     5.38 &     0.02 &    -0.03\\
PG 1425+267    &     7.85 &     5.78 &     7.92 &     5.74 &     0.07 &    -0.05\\
PG 1427+480    &     8.21 &     5.80 &     8.23 &     5.76 &     0.02 &    -0.04\\
PG 1440+356    &     6.30 &     4.28 &     6.31 &     4.19 &     0.01 &    -0.09\\
PG 1444+407    &     7.33 &     5.39 &     7.32 &     5.35 &    -0.00 &    -0.04\\
PG 1512+370    &     8.17 &     6.14 &     8.24 &     6.09 &     0.07 &    -0.05\\
PG 1543+489    &     7.04 &     4.91 &     7.12 &     4.82 &     0.08 &    -0.08\\
PG 1545+210    &     7.99 &     5.94 &     7.99 &     5.96 &     0.00 &     0.02\\
PG 1626+554    &     8.22 &     6.75 &     8.06 &     6.66 &    -0.16 &    -0.08\\
PG 1704+608    &     7.08 &     4.83 &     7.15 &     4.83 &     0.07 &    -0.00\\
\hline
Mean \footnotemark[4]& & & & &    -0.01 &    -0.02 \\
Standard Deviation & & & & &     0.08 &     0.06 \\
\hline
\end{tabular}
\footnotetext[1]{The WISE W3 and W4 bands are at 12 and 22 $\mu$m, respectively.}
\footnotetext[2]{The change in magnitude is taken as the difference in the WISE and {\it Spitzer} magnitudes, specifically, WISE$_{\textrm{All-Sky}}$-WISE$_{{\it Spitzer}}$.}
\footnotetext[3]{Large values of $\Delta$W3 and $\Delta$W4 for this object indicate significant variation of infrared emission and a likely synchrotron component.}
\footnotetext[4]{4C 11.69 is excluded when computing these statistics.}
\end{minipage}
\end{table*}

%% file: sample.tex
\begin{table*}
\begin{minipage}[2cm]{18.5cm}
\caption{Sample data \label{tab:sample}}
\renewcommand{\thefootnote}{\alph{footnote}}
\begin{tabular}{lccrrrrrrr}
Object & Redshift & log$(L_{\textrm{iso}})$ & $\zeta_{\textrm{ratio},1.5\mu\textrm{m}}$ & $\zeta_{\textrm{ratio},2\mu\textrm{m}}$ & $\zeta_{\textrm{ratio},3\mu\textrm{m}}$ & $\zeta_{\textrm{ratio},7\mu\textrm{m}}$ & $\zeta_{\textrm{ratio},12\mu\textrm{m}}$ & $\zeta_{\textrm{ratio},15\mu\textrm{m}}$ & $\zeta_{\textrm{ratio},24\mu\textrm{m}}$ \\
       & 	    & ergs s$^{-1}$ & 				  &			     &	 			&		&	&			\\
\hline
3C 215    &      0.4108 & 45.77 &    11.03& \nodata & \nodata & \nodata & \nodata & \nodata & \nodata \\
3C 232    &      0.5297 & 46.36 &     7.21& \nodata & \nodata & \nodata & \nodata & \nodata & \nodata \\
3C 254    &      0.7363 & 46.17 &     4.97 &     5.36 &     4.66 &     4.46 &     3.71 &     3.76 &     6.02\\
3C 263    &      0.6464 & 46.81 &    10.85 &    10.11 &     9.20 &    10.29 &     9.41 &     9.76 &    10.52\\
3C 277.1    &      0.3199 & 45.61 &    14.76& \nodata & \nodata & \nodata & \nodata & \nodata & \nodata \\
3C 281    &      0.6017 & 46.23 &    12.81& \nodata & \nodata & \nodata & \nodata & \nodata & \nodata \\
3C 334    &      0.5553 & 46.44 &    10.50 &    10.32 &     9.99 &     9.49 &     5.79 &     5.16 &     4.61\\
3C 37    &      0.6661 & 46.18 &     8.96& \nodata & \nodata & \nodata & \nodata & \nodata & \nodata \\
3C 446    &      1.4040 & 47.01 &     1.64& \nodata & \nodata & \nodata & \nodata & \nodata & \nodata \\
3C 47    &      0.4250 & 45.97 &    10.32 &     7.28 &     4.41 &     3.59 &     3.32 &     3.32 &     4.31\\
4C 01.04    &      0.2634 & 45.44 &     5.39& \nodata & \nodata & \nodata & \nodata & \nodata & \nodata \\
4C 06.69    &      1.0002 & 47.30 &     9.12& \nodata & \nodata & \nodata & \nodata & \nodata & \nodata \\
4C 10.06    &      0.4075 & 46.48 &    18.53& \nodata & \nodata & \nodata & \nodata & \nodata & \nodata \\
4C 19.44    &      0.7192 & 46.91 &    13.30& \nodata & \nodata & \nodata & \nodata & \nodata & \nodata \\
4C 20.24    &      1.1135 & 46.92 &    10.79& \nodata & \nodata & \nodata & \nodata & \nodata & \nodata \\
4C 22.26    &      0.9760 & 46.59 &    12.85& \nodata & \nodata & \nodata & \nodata & \nodata & \nodata \\
4C 31.63    &      0.2952 & 46.61 &    15.74 &    13.04 &    11.71 &    12.39 &    12.75 &    13.58 &    21.09\\
4C 34.47    &      0.2055 & 46.08 &    17.16 &    15.05 &    14.14 &    14.52 &    12.45 &    12.89 &    21.26\\
4C 39.25    &      0.6946 & 46.91 &    15.66& \nodata & \nodata & \nodata & \nodata & \nodata & \nodata \\
4C 40.24    &      1.2520 & 46.60 &     6.08& \nodata & \nodata & \nodata & \nodata & \nodata & \nodata \\
4C 41.21    &      0.6124 & 46.75 &    14.97& \nodata & \nodata & \nodata & \nodata & \nodata & \nodata \\
4C 49.22    &      0.3333 & 45.99 &    10.68& \nodata & \nodata & \nodata & \nodata & \nodata & \nodata \\
4C 55.17    &      0.8990 & 46.46 &     3.65& \nodata & \nodata & \nodata & \nodata & \nodata & \nodata \\
4C 58.29    &      1.3740 & 47.20 &    59.70& \nodata & \nodata & \nodata & \nodata & \nodata & \nodata \\
4C 73.18    &      0.3027 & 46.40 &     9.78& \nodata & \nodata & \nodata & \nodata & \nodata & \nodata \\
B2 0742+31    &      0.4616 & 46.46 &     5.94& \nodata & \nodata & \nodata & \nodata & \nodata & \nodata \\
IRAS F07546+3928    &      0.0953 & 45.43 &     8.60 &     4.48 &     4.03 &     4.69 &     4.44 &     3.80 &     4.07\\
MRK 509    &      0.0345 & 45.36 &    12.74 &    10.90 &    11.73 &    12.63 &    11.84 &    10.91 &    11.36\\
OS 562    &      0.7506 & 46.74 &     5.84& \nodata & \nodata & \nodata & \nodata & \nodata & \nodata \\
PG 0052+251    &      0.1544 & 45.94 &    15.10 &    12.72 &    13.38 &    13.39 &    13.20 &    12.44 &    21.14\\
PG 0844+349    &      0.0643 & 45.31 &    28.05 &    21.11 &    17.48 &    17.89 &    15.31 &    17.73 &    21.25\\
PG 0947+396    &      0.2057 & 45.79 &     9.99 &     7.20 &     6.66 &     6.90 &     7.59 &     8.54 &     9.98\\
PG 0953+414    &      0.2338 & 46.42 &    24.26 &    19.44 &    15.99 &    22.19 &    21.68 &    25.69 &    39.96\\
PG 1001+054    &      0.1603 & 45.13 &     9.93 &     4.58 &     3.26 &     3.00 &     3.58 &     4.02 &     4.86\\
PG 1100+772    &      0.3114 & 46.46 &    18.90 &    16.32 &    12.72 &    15.70 &    13.49 &    15.59 &    19.96\\
PG 1114+445    &      0.1440 & 45.39 &     9.54 &     4.90 &     3.74 &     2.85 &     2.53 &     2.65 &     3.15\\
PG 1115+407    &      0.1541 & 45.72 &    22.00 &    13.84 &    12.55 &    13.21 &    13.65 &    14.92 &    16.69\\
PG 1116+215    &      0.1759 & 46.31 &    18.55 &    11.49 &     8.89 &    11.80 &    13.54 &    15.85 &    19.19\\
PG 1202+281    &      0.1651 & 45.41 &     9.92 &     6.51 &     5.58 &     5.29 &     3.52 &     3.71 &     3.82\\
PG 1216+069    &      0.3319 & 46.31 &    14.55 &    14.32 &    12.09 &    12.37 &    15.71 &    17.43 &    36.00\\
PG 1226+023    &      0.1576 & 46.96 &    19.03 &    14.46 &    12.75 &    14.15 &    16.76 &    16.87 &    19.03\\
PG 1309+355    &      0.1823 & 45.74 &    11.19 &     8.45 &     6.72 &     7.77 &     4.45 &     4.45 &     4.93\\
PG 1322+659    &      0.1684 & 45.74 &    15.71 &    10.93 &     9.56 &    10.58 &    11.35 &    12.98 &    13.64\\
PG 1351+640    &      0.0882 & 45.31 &    12.39 &     7.91 &     6.12 &     4.65 &     2.54 &     2.58 &     1.98\\
PG 1352+183    &      0.1510 & 45.61 &    29.16 &    16.96 &    20.23 &    19.64 &    16.76 &    13.94 &    19.38\\
PG 1402+261    &      0.1650 & 46.02 &    18.67 &    11.01 &     8.63 &     9.14 &     8.97 &    10.03 &    10.88\\
PG 1411+442    &      0.0895 & 45.33 &     9.46 &     5.63 &     4.73 &     3.94 &     5.02 &     5.75 &     7.94\\
PG 1415+451    &      0.1143 & 45.37 &     9.84 &     7.47 &     8.25 &     9.03 &     8.56 &     8.94 &     9.66\\
PG 1425+267    &      0.3637 & 46.05 &     8.51 &     7.26 &     5.70 &     4.97 &     4.50 &     4.51 &     4.89\\
PG 1427+480    &      0.2203 & 45.79 &    18.26 &    12.95 &    13.23 &    11.86 &    10.62 &     9.47 &     7.55\\
PG 1440+356    &      0.0773 & 45.64 &    10.21 &     8.46 &     9.74 &    12.41 &    14.08 &    15.32 &    13.48\\
PG 1444+407    &      0.2673 & 46.20 &    15.67 &    12.55 &    10.47 &     9.81 &     8.27 &     9.51 &    10.22\\
PG 1512+370    &      0.3700 & 46.22 &    12.68 &    10.91 &     9.65 &    10.49 &     9.12 &     9.32 &    10.75\\
PG 1543+489    &      0.4000 & 46.30 &     7.44 &     6.31 &     5.18 &     3.71 &     2.99 &     2.94 &     2.75\\
PG 1545+210    &      0.2642 & 45.93 &    12.24 &     8.47 &     6.44 &     8.32 &     7.54 &     8.82 &    14.58\\
PG 1626+554    &      0.1317 & 45.68 &    27.38 &    16.00 &    15.59 &    23.00 &    27.67 &    38.40 &    71.14\\
PG 1704+608    &      0.3730 & 46.49 &     7.80 &     6.39 &     5.12 &     7.47 &     5.99 &     5.72 &     4.96\\
PG 2349−014    &      0.1740 & 45.77 &    13.63 &     6.60 &     6.25 &     7.55 &     7.30 &     7.73 &     6.92\\
PKS 0112-017    &      1.3743 & 46.88 &    38.31& \nodata & \nodata & \nodata & \nodata & \nodata & \nodata \\
PKS 0403-13    &      0.5700 & 46.36 &     8.52& \nodata & \nodata & \nodata & \nodata & \nodata & \nodata \\
PKS 1127-14    &      1.1870 & 47.15 &     4.57& \nodata & \nodata & \nodata & \nodata & \nodata & \nodata \\
PKS 1656+053    &      0.8890 & 47.06 &    13.74& \nodata & \nodata & \nodata & \nodata & \nodata & \nodata \\
\hline
\end{tabular}
\end{minipage}
\end{table*}

%% file: stats.tex
\begin{table}
\begin{minipage}[2cm]{14.5cm}
\caption{Bolometric ratio statistics \label{tab:stats}}
\begin{tabular}{rrrr}
Wavelength & Median & Mean & Std. Dev. \\
\hline
1.5 $\mu$m &     12.2 &     13.8 &      8.9 \\
2 $\mu$m &     10.3 &     10.5 &      4.3 \\
3 $\mu$m &      9.2 &      9.4 &      4.2 \\
7 $\mu$m &      9.8 &     10.1 &      5.2 \\
12 $\mu$m &      9.0 &      9.7 &      5.7 \\
15 $\mu$m &      9.5 &     10.5 &      7.2 \\
24 $\mu$m &     10.5 &     13.9 &     13.1 \\
\hline
\end{tabular}
\end{minipage}
\end{table}

%% file: corrections.tex
\begin{table*}
\begin{minipage}[2cm]{18cm}
\caption{Infrared Bolometric Corrections \label{tab:fits}}
\renewcommand{\thefootnote}{\alph{footnote}}
\begin{tabular}{rcccc}
Wavelength & Bolometric correction with zero intercept & Bolometric correction with nonzero intercept & Sig. of nonzero intercept \footnotemark[1] \\
\hline
1.5 $\mu$m & $L_{\textrm{iso}}=(    11.8\pm     0.9)\,\lambda L_{\lambda}$ & log$(L_{\textrm{iso}})=(    8.98\pm    2.03)\,+\,(    0.82\pm    0.05)\,\textrm{log}(\lambda L_{\lambda})$ & 4.42 (100.0\%)\\
2 $\mu$m & $L_{\textrm{iso}}=(     9.6\pm     0.7)\,\lambda L_{\lambda}$ & log$(L_{\textrm{iso}})=(    1.85\pm    3.18)\,+\,(    0.98\pm    0.07)\,\textrm{log}(\lambda L_{\lambda})$ & 0.58 ($<$44.8\%) \\
3 $\mu$m & $L_{\textrm{iso}}=(     8.4\pm     0.7)\,\lambda L_{\lambda}$ & log$(L_{\textrm{iso}})=(    4.54\pm    3.42)\,+\,(    0.92\pm    0.08)\,\textrm{log}(\lambda L_{\lambda})$ & 1.29 (79.8.0\%) \\
7 $\mu$m & $L_{\textrm{iso}}=(     8.8\pm     0.8)\,\lambda L_{\lambda}$ & log$(L_{\textrm{iso}})=(    1.85\pm    4.03)\,+\,(    0.88\pm    0.09)\,\textrm{log}(\lambda L_{\lambda})$ & 1.57 (88.1\%) \\
12 $\mu$m & $L_{\textrm{iso}}=(     8.1\pm     0.9)\,\lambda L_{\lambda}$ & log$(L_{\textrm{iso}})=(    8.92\pm    4.30)\,+\,(    0.82\pm    0.10)\,\textrm{log}(\lambda L_{\lambda})$ & 2.08 (95.7\%) \\
15 $\mu$m & $L_{\textrm{iso}}=(     8.5\pm     1.0)\,\lambda L_{\lambda}$ & log$(L_{\textrm{iso}})=(   10.51\pm    4.39)\,+\,(    0.79\pm    0.10)\,\textrm{log}(\lambda L_{\lambda})$ & 2.38 (97.8\%)\\
24 $\mu$m & $L_{\textrm{iso}}=(    10.1\pm     1.4)\,\lambda L_{\lambda}$ & log$(L_{\textrm{iso}})=(   15.03\pm    4.77)\,+\,(    0.69\pm    0.11)\,\textrm{log}(\lambda L_{\lambda})$ & 3.15 (99.6\%) \\
\hline
\end{tabular}
\footnotetext[1]{The approximate significance of adding a nonzero intercept is given by the t-ratio with the approximate associated probability given in parentheses.}
\end{minipage}
\end{table*}

%% file: fit1.tex
\begin{eqnarray}
\label{eqn:fit1}
\textrm{log}(L_{\textrm{iso}})=(    8.98\pm    2.03)\,+\,(    0.82\pm    0.05)\,\textrm{log}(\lambda L_{\lambda})
\end{eqnarray}

%% file: fit2.tex
\begin{eqnarray}
\label{eqn:fit2}
\textrm{log}(L_{\textrm{iso}})=(    1.85\pm    3.18)\,+\,(    0.98\pm    0.07)\,\textrm{log}(\lambda L_{\lambda})
\end{eqnarray}

%% file: fit3.tex
\begin{eqnarray}
\label{eqn:fit3}
\textrm{log}(L_{\textrm{iso}})=(    4.54\pm    3.42)\,+\,(    0.92\pm    0.08)\,\textrm{log}(\lambda L_{\lambda})
\end{eqnarray}

%% file: fit7.tex
\begin{eqnarray}
\label{eqn:fit7}
\textrm{log}(L_{\textrm{iso}})=(    6.31\pm    4.03)\,+\,(    0.88\pm    0.09)\,\textrm{log}(\lambda L_{\lambda})
\end{eqnarray}

%% file: fit12.tex
\begin{eqnarray}
\label{eqn:fit12}
\textrm{log}(L_{\textrm{iso}})=(    8.92\pm    4.30)\,+\,(    0.82\pm    0.10)\,\textrm{log}(\lambda L_{\lambda})
\end{eqnarray}

%% file: fit15.tex
\begin{eqnarray}
\label{eqn:fit15}
\textrm{log}(L_{\textrm{iso}})=(   10.51\pm    4.39)\,+\,(    0.79\pm    0.10)\,\textrm{log}(\lambda L_{\lambda})
\end{eqnarray}

%% file: fit24.tex
\begin{eqnarray}
\label{eqn:fit24}
\textrm{log}(L_{\textrm{iso}})=(   15.03\pm    4.77)\,+\,(    0.69\pm    0.11)\,\textrm{log}(\lambda L_{\lambda})
\end{eqnarray}